\documentclass[a4paper,preprintnumbers,floatfix,superscriptaddress,prl,twocolumn,showpacs,notitlepage,longbibliography]{revtex4-1}
\usepackage[utf8]{inputenc}
\usepackage[T1]{fontenc}
\usepackage[sc,osf]{mathpazo}
\usepackage{amsmath, amsthm, amssymb,amsfonts,mathbbol,amstext}
\usepackage{graphicx}
\usepackage{dcolumn}
\usepackage{bm}
\usepackage{bbm}
\usepackage{hyperref}
\usepackage{mathtools}
\usepackage{comment}
\usepackage{color}
\usepackage{url}
\usepackage{multirow}
\usepackage{diagbox}

\definecolor{applegreen}{rgb}{0.55, 0.71, 0.0}
\definecolor{darkelectricblue}{rgb}{0.03, 0.51, 0.57}

\newcommand{\ket}[1]{| #1 \rangle}
\newcommand{\bra}[1]{\langle #1 |}

\def\RR{\mathbbm{R}}

\def\1{\mathbf{1}}
\def\0{\mathbf{0}}

\def\p{\mathbf{p}}

\def\p{\mathbf{p}}

\begin{document}

\title{Device-Independent Secret Sharing and a Stronger Form of Bell non-locality}
\author{M. G. M. Moreno}
\affiliation{International Institute of Physics, Federal University of Rio Grande do Norte, 59070-405 Natal, Brazil}
\author{Samura\'i Brito}
\affiliation{International Institute of Physics, Federal University of Rio Grande do Norte, 59070-405 Natal, Brazil}
\author{Ranieri V. Nery}
\affiliation{International Institute of Physics, Federal University of Rio Grande do Norte, 59070-405 Natal, Brazil}
\author{Rafael Chaves}
\affiliation{International Institute of Physics, Federal University of Rio Grande do Norte, 59070-405 Natal, Brazil}
\affiliation{School of Science and Technology, Federal University of Rio Grande do Norte, 59078-970 Natal, Brazil}

\date{\today}
\begin{abstract}
Bell non-locality, the fact that local hidden variable models cannot reproduce the correlations obtained by measurements on entangled states, is a cornerstone in our modern understanding of quantum theory. Apart from its fundamental implications, non-locality is also at the core of device-independent quantum information processing, which successful implementation is achieved without precise knowledge of the physical apparatus. Here we show that a stronger form of Bell non-locality, for which even non-local hidden variable models cannot reproduce the quantum predictions, allows for the device-independent implementation of secret sharing, a paradigmatic communication protocol where a secret split amidst many possibly untrusted parts can only be decoded if they collaborate among themselves.
\end{abstract}
\maketitle

Among the many promises of quantum technologies, quantum communication \cite{gisin2007quantum} is arguably at most advanced stage to break out of the lab and reach practical large scale use, most prominently in quantum cryptography \cite{gisin2002quantum} that, indeed, is already commercially available. Recent breakthroughs, such as the first loophole free violation of Bell inequalities \cite{Hensen2015,Giustina2015,Shalm2015} and the launch of a satellite capable of distributing quantum entanglement across very large distances \cite{Yin2017}, made the goal of scalable quantum networks \cite{Kimble2008} exchanging information in a fundamentally secure way \cite{Ekert1991}, even more tangible.

Central to quantum cryptography --for instance in the famous BB84 protocol \cite{bennett2014quantum}-- is the intrinsic randomness of quantum mechanics, the fact that the outcomes of measurements performed on quantum systems cannot (in most cases) be predicted with certainty. Such protocols, unfortunately, have to rely on a precise characterization and trust of measurements devices, otherwise they can and in fact have already been hacked \cite{lydersen2010hacking}. Hacking, however, becomes impossible in a broader framework known as device-independent (DI) quantum information processing \cite{mayers1998quantum,PhysRevLett.95.010503,PhysRevLett.113.140501}. At the center of many DI applications is the phenomenon known as Bell non-locality \cite{Bell1964}, the fact the correlations obtained by local measurements on distant entangled states are incompatible with any local hidden variable (LHV) model. Independently of the precise knowledge about the quantum states being prepared or which measurements are being performed, the violation of a Bell inequality puts strict bounds on the amount of information that any eavesdropper can have access to \cite{Colbeck2007,Pironio2010}, providing the ultimate cryptographic security that cannot be surpassed unless the very laws of physics are broken. 

In parallel to that, in recent years, LHV models have been recognized as a particular and simple case of causal Bayesian networks \cite{Pearl2009,Fritz2014,Spekkens2015}, giving rise to new and stronger forms of non-classicality \cite{Chaves2015b,renou2019genuine} as compared to the paradigmatic Bell non-locality.  And, since non-locality is a resource in a variety of DI applications \cite{Brunner2014}, it is natural to ask if these novel forms of non-classical behavior can also be put to use in information processing. As it turns out, the non-locality in quantum networks composed by independent sources also limits the information of an eavesdropper \cite{lee2018towards} and temporal networks reminiscent of the instrumental causal structure \cite{Chaves2018} provide a new and more efficient platform for randomness certification \cite{agresti2019experimental}. In spite of all progress, however, one paradigmatic quantum communication protocol has yet not found a DI formulation: the secret sharing \cite{PhysRevA.59.1829,PhysRevA.59.162}. In a secret sharing protocol, a secret is split into $n - 1$ parts and sent to different receivers in a way that at least $m\leq n-1$ of them must collaborate to reveal it. Even though, initial attempts \cite{PhysRevLett.108.100401} indicated the possibility of a DI certification of secret sharing, the fact that some of the receivers might be untrusted introduces a new class of possible attacks \cite{woodhead2018randomness} that are not covered by the analysis based on the standard Bell non-locality. 

In this paper, we show that the DI certification of a $n$-partite secret sharing protocol with up to $n-2$ untrusted parties can be achieved via the violation of Bell inequalities witnessing a stronger than Bell type of non-locality. We employ the causal approach in \cite{chaves2017causal} to show that the attack by an untrusted receiver can be mapped to relaxations of the locality assumption in Bell's theorem \cite{Chaves2015b} and that can be characterized by the so-called Svetlichny inequality \cite{Svet1987}. By considering quantum as well as non-signaling (NS) resources \cite{Popescu1994}, we first analytically show that maximum violation of this inequality guarantees full security. Further, by writing the corresponding optimization problem as a semi-definite program we also show that the higher is the violation of the Svetlichny inequality, the smaller is the information that an untrusted receiver as well as an external eavesdropper can get about the secret to be shared. 

\begin{figure}[!t]
    \centering
    \includegraphics[scale=0.32]{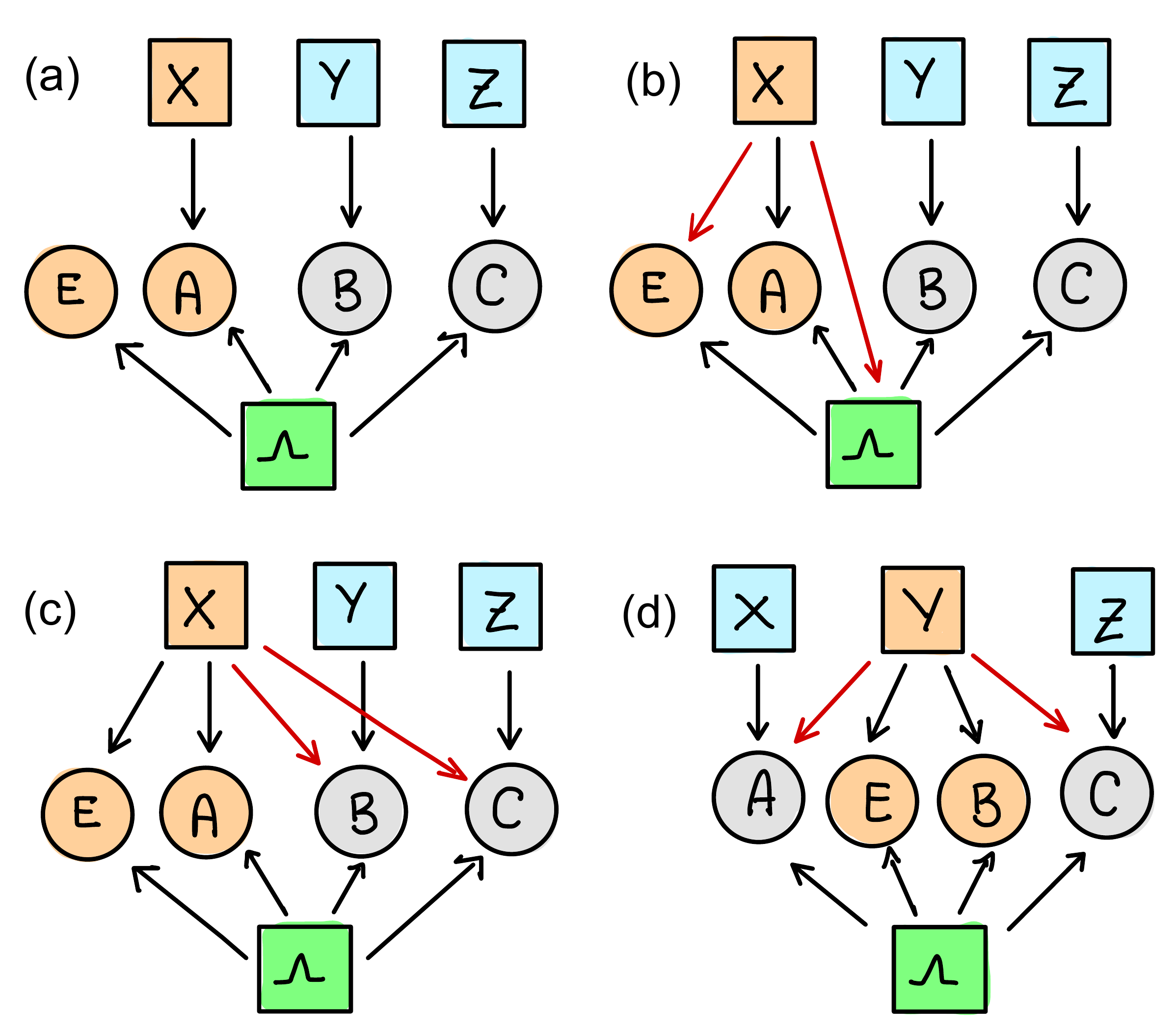}
    \caption{\textbf{DAG representation of causal structures.} \textbf{a)} Tripartite scenario with an external eavesdropper that can share correlations with Alice, Bob and Charlie but has no direct access to their inputs or ouputs. \textbf{b)} If Alice is untrusted, she can correlate her input with the source $\Lambda$ and try to guess the output $c$ via an extra outcome $e$ (unknown to Bob and Charlie). \textbf{c)} Causal structure where the input of Alice is broadcast to Bob and Charlie and can generate the same NS correlations as the DAG in Fig. \ref{fig:causals}b). \textbf{d)} DAG characterizing the correlations obtainable by an untrusted Bob.}
    \label{fig:causals}
\end{figure}

\emph{Secret sharing, causal models and the attack by an untrusted receiver-- } Without loss of generality, we will focus here in a tripartite secret sharing protocol between three distant parties: Alice and Bob (the receivers) and Charlie (the sender). Each of them have parts of a correlated device (see Fig. \ref{fig:causals}) which receives binary inputs ($x=0,1$, $y=0,1$ and $z=0,1$ for Alice, Bob and Charlie respectively) and returns binary outputs ($a=0,1$, $b=0,1$ and $c=0,1$, respectively). The device is thus described by a probability distribution $p(a,b,c \vert x,y,z)$. Charlie's aim is to share a secret bit $c^*$ with Alice and Bob, in such a way that the their individual bits ($a^*$ and $b^*$, respectively) are random and have no information about $c^*$, for instance
\begin{eqnarray}
c^*=a^*\oplus b^*,
\end{eqnarray}
where $\oplus$ represents sum mod(2) and $a^*$, $b^*$ and $c^*$ are the measurement outputs for some specific inputs $x=x^*,y=y^*,z=z^*$. That is, Alice and Bob have to collaborate in order to retrieve any information about $c^*$. Furthermore, this should be achieved in a secure way against a fourth malicious part (an eavesdropper) and also assuming that either Alice or Bob might be untrusted (try to recover $c^*$ without the help of the other). As proposed in \cite{PhysRevA.59.1829,PhysRevA.59.162}, it is possible to achieve secret sharing against an eavesdropper if they share a quantum state and each measure observables $A_x$, $B_y$ and $C_z$ such that the
\begin{equation}
p(a,b,c \vert x,y,z)= \mathrm{Tr} \left[ \left( M^a_x \otimes M^b_y \otimes M^c_z \right) \rho  \right]  
\end{equation}
where $A_x=M^0_x-M^1_x$ and $M^a_x$ is a POVM operator (similarly for Bob and Charlie), $A_0=B_0=C_0=\sigma_x$ and $A_1=B_1=C_1=\sigma_y$ ($\sigma_x$ and $\sigma_y$ being the Pauli matrices) and $\rho=\ket{GHZ}\bra{GHZ}$ is the GHZ state \cite{greenberger1989going} with $\ket{GHZ}= 1/\sqrt{2}\left( \ket{000}+\ket{111} \right)$. For instance, if $x^*=y^*=z^*=0$ it follows that $a^* \oplus b^* = c^*$, thus fulfilling the condition for secret sharing. The DI security against an eavesdropper can be seen from the fact that the obtained distribution $p(a,b,c \vert x,y,z)$ maximally violates the Mermin inequality \cite{Mermin1990} and such violation can only be achieved if the underlying quantum state shared between Alice, Bob and Charlie is a pure GHZ state \cite{colbeck2009quantum,colbeck2011private,PhysRevLett.117.070402}, thus guaranteeing that any fourth party (the eavesdropper) has no correlations whatsoever with them (and thus no information about their measurement outcomes and the secret).

But what if one of the receivers, Alice or Bob, are untrusted? To illustrate, suppose Alice is the untrusted part (Fig. \ref{fig:causals}b). Her aim is to simulate a non-local correlation satisfying the secret sharing condition $a^* \oplus b^* = c^*$ but, at the same time, generate an extra output $e$ conveying the most information about the secret bit $c^*$. That is, she wants to maximize $p(e=c\vert x^*y^*z^*)$, without the need of collaboration with Bob. We can model this scenario by a joint probability $p(a,b,c,e\vert x,y,z)$ that marginalizes to the observed distribution in the execution of the protocol, that is, $\sum_{e}p(a,b,c,e\vert x,y,z)=p(a,b,c\vert x,y,z)$. As recently pointed out \cite{woodhead2018randomness}, if the eavesdropper is among the receivers, a new class of attack becomes possible. In order to model  it we represent the underlying causal structure as a directed acyclic graph (DAG) \cite{Pearl2009} as shown in Fig. \ref{fig:causals}. The vertices denote the variables of interest (inputs, outputs and source of correlations) and the direct edges encode their causal relations. As paradigmatic in any DI cryptographic protocol \cite{PhysRevLett.95.010503,PhysRevLett.113.140501}, an external eavesdropper has no direct access to the variables of the other parties, the most it can do is to share correlations with the others, for instance, having part of the joint quantum state. This is captured by the DAG in Fig. \ref{fig:causals}a, where the eavesdropper outcome $e$ is only allowed to have causal dependence on the source of correlations $\lambda$. However, if the eavesdropper is one of the receivers, it can now generate correlations between its own input and the source of correlations (see Fig. \ref{fig:causals}b). In the language of Bell's theorem, this would be equivalent to a measurement dependent hidden variable model \cite{hall2016significance,big2018challenging,Friedman2019}; in other terms, we cannot assume ``free-will'', as the measurement to be performed by the untrusted part depends now on the common source of correlations. Within this new causal structure, it is easy to devise hidden variable models mimicking the violation of Mermin's inequality (see Appendix for further details). That is, Mermin's violation is no longer enough to guarantee the DI security of the protocol.

\emph{Stronger forms of non-locality and DI secret sharing-- }The key insight behind our approach is that to be secure against the generalized model in Fig. \eqref{fig:causals}b one first should consider the Bell inequalities characterizing the corresponding DAG. And, as we do not know which of the two receivers, Alice or Bob, is the untrusted part, this should also be taken into account. As detailed in the Appendix, the main ingredient for the derivation of Bell inequalities in this scenario is the fact that all correlations that can be generated by the measurement dependent model in Fig. \ref{fig:causals}b can also be generated by the non-local HV model in Fig. \ref{fig:causals}c. Following this prescription, we find that only one class of full-correlator Bell inequalities characterize this scenario (also considering the symmetry in Fig. \ref{fig:causals}d in which Bob is the untrusted part), the well known Svetlichny inequality \cite{Svet1987}, that we rewrite here as
\begin{eqnarray}\label{snew}
S & & = p_A(0 \vert 0) CHSH_{00}-p_A(1 \vert 0) CHSH_{10} \\ \nonumber
& & + p_A(0 \vert 1) CHSH^{\prime}_{01}-p_A(1 \vert 1) CHSH^{\prime}_{11} \leq  4,
\end{eqnarray}
where $p_A(a \vert x)$ is the marginal probability of Alice; $CHSH_{ax}$ and $CHSH^{\prime}_{ax}$ refer to symmetries of the CHSH inequality \cite{Clauser1969} given by $CHSH_{ax}=E^{ax}_{00}+E^{ax}_{01}+E^{ax}_{10}-E^{ax}_{11}$ and $CHSH^{\prime}_{ax}=E^{ax}_{00}-E^{ax}_{01}-E^{ax}_{10}-E^{ax}_{11}$ and  $E^{ax}_{yz}=\sum_{b,c=0,1} (-1)^{b+c}p(b,c \vert x,a,y,z)$ is the expectation value of the measurement outcome of Bob and Charlie conditioned on a given outcome $a$ and the input $x$ of Alice. Clearly, because of the normalization of $p_A(a \vert x)$, to achieve the maximum quantum violation $S=4\sqrt{2}$ ($S=8$ for the NS set) necessarily we must have $CHSH_{10}=CHSH^{\prime}_{11}=-2\sqrt{2}$ and $CHSH_{00}=CHSH^{\prime}_{01}=2\sqrt{2}$ (or $CHSH_{10}=CHSH^{\prime}_{11}=-4$ and $CHSH_{00}=CHSH^{\prime}_{01}=4$ in the NS scenario). That is, independently of the measurement outcome of Alice and her input (which can affect both outcomes $b$ and $c$, considering that she is the untrusted party), the conditional statistics of Bob and Charlie have to maximally violate the CHSH inequality. This is known to be possible if and only if the state of Bob and Charlie (conditioned on $a$ and $x$) is the maximally entangled state of two qubits \cite{mayers1998quantum,mayers2003self} for the quantum set, and a PR-box in the non-signaling set. Thus, if a maximum quantum violation of the Svetlichny inequality is observed, Bob and Charlie can be sure that their conditioned state is uncorrelated with any other system, that is, if we are restricted to quantum correlations we have that $\rho_{BCE}=\rho_{BC}\otimes \rho_E$. The same line of reasoning follows if Bob is the untrusted part. Hence, $p(e=c\vert x^*y^*z^*)=1/2$, thus guaranteeing the full security in the secret sharing protocol. We once more highlight that a self-testing result based on the Mermin inequality is not enough in a scenario with an untrusted part, as its violation could be simulated by purely classical resources. On contrast, since the Svetlichny violation witness a stronger form of non-locality, even an untrusted part is not able to fake it. The same argument can be employed to certify security in a $n$-partite scenario, one sender and $n-1$ receivers, such that the secret is safe even against an attack of $n-2$ dishonest parts which may collaborate with each other, see Appendix for details. The result above shows that stronger forms of nonlocality can lead to new DI protocols. However, as stated, it requires maximum Svetlichny violation, that in practice cannot be achieved. Nicely, as we show next, the result is robust, displaying security even far from maximum violation.  

The joint probability $p(a,b,c,e|x,y,z)$ should respect the causal constraints imposed by the causal model underlying an untrusted receiver. For instance, if Alice is untrusted (Fig. \ref{fig:causals}c), in generalized probabilistic framework \cite{Popescu1994} this equals to impose the following constraints (plus the normalization and positivity conditions):
\begin{eqnarray}
\label{NSconditions}
\nonumber
&\sum_b p(a,b,c,e|x,y,z)=\sum_b p(a,b,c,e|x,y^{\prime},z),&\\
&\sum_c p(a,b,c,e|x,y,z)=\sum_c p(a,b,c,e|x,y,z^{\prime})&,
\end{eqnarray}
referring to the non-signaling (NS) condition for Bob and Charlie, respectively. It should also be noted that the scenario of Fig. \ref{fig:causals}c allows for a marginal distribution with signalling from Alice to Bob and Charlie. However, for her attack to be successful and not detectable by the other parts, the observable tripartite distribution $p(abc\vert xyz)$ must also be non-signaling from Alice to the other parties, that is
\begin{eqnarray*}
&\sum_{a,e} p(abce\vert xyz)=\sum_{a,e} p(abce\vert x^{\prime}yz).
\end{eqnarray*}
\begin{figure}[!t]
    \centering
    \includegraphics[scale=0.5]{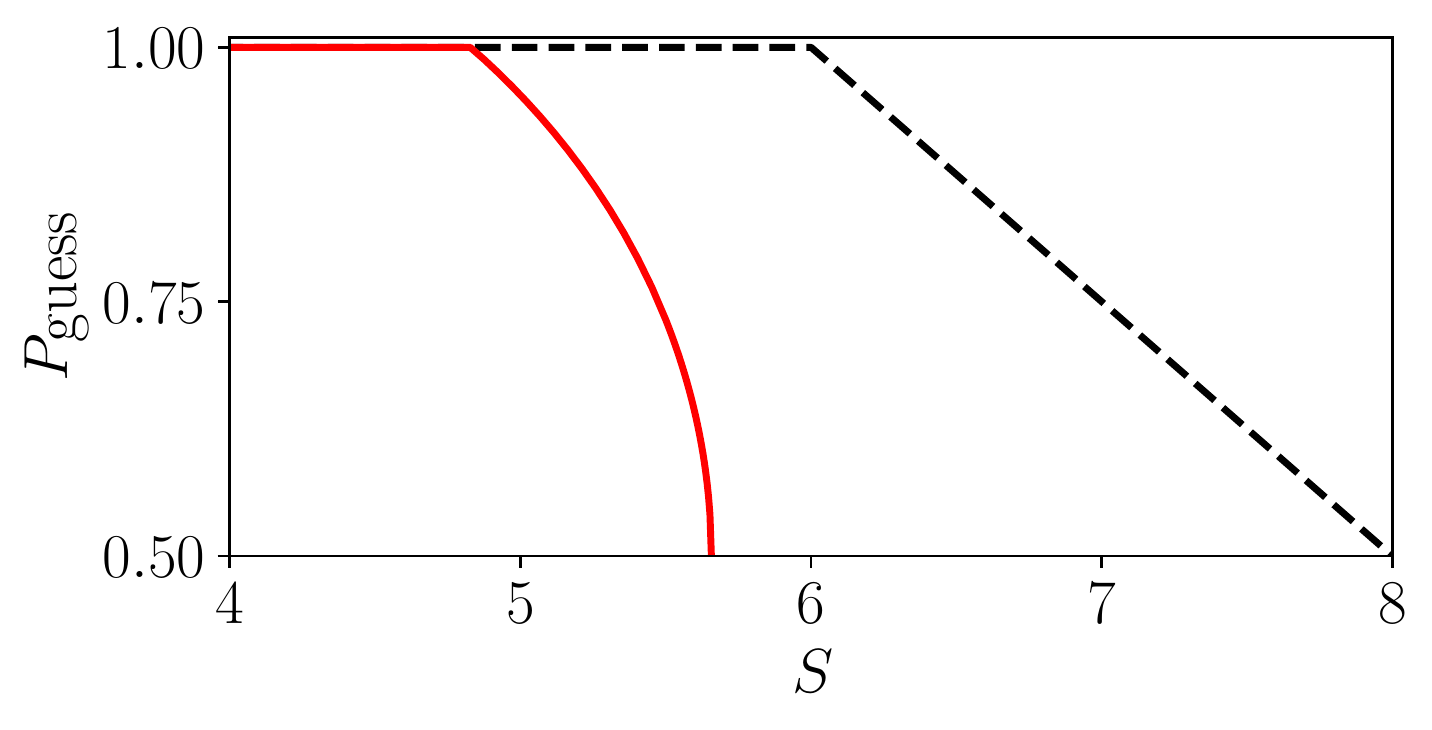}
    \caption{\textbf{Guessing probability as a function of Svetlichny's inequality violation.} Upper bound for Eve's guessing probability $P_{\textrm{guess}} \coloneqq p(e=c\vert x^*y^*z^*)$ for a generic quantum behaviour is shown as the solid curve. This bound is obtained from the level 2+ABC+ABE+BCE+ABCE of the NPA hiearchy \cite{NPA1}. Even with the approximation, it can be seen that for the maximal quantum violation $4\sqrt{2}$ of the inequality, perfect security for Charlie's bit can be obtained as the guessing probability approaches $0.5$. The dashed curve represents the optimal NS guessing probability.}
    \label{fig:plot1}
\end{figure}

Alternatively, in a quantum description we have that
\begin{equation}
\label{qconditions}
p(abce|xyz)= \mathrm{Tr} \left[ \left(M^e_x \otimes M^a_x \otimes M^b_{x,y} \otimes M^c_{x,z} \right) \rho_{ABCE}  \right],
\end{equation}
and that in practice can be casted as an hierarchy of approximations \cite{NPA1} implemented as semi-definite programs. Similar expressions follow to the situation where Bob is the untrusted part (Fig. \ref{fig:causals}d). We also highlight, that as the DAGs in Fig. \ref{fig:causals}b-d have the usual Bell DAG (Fig. \ref{fig:causals}a) as a particular case, all the results derived for an untrusted part also clearly will hold for an external eavesdropper.

\begin{figure}[!t]
    \centering
    \includegraphics[scale=0.5]{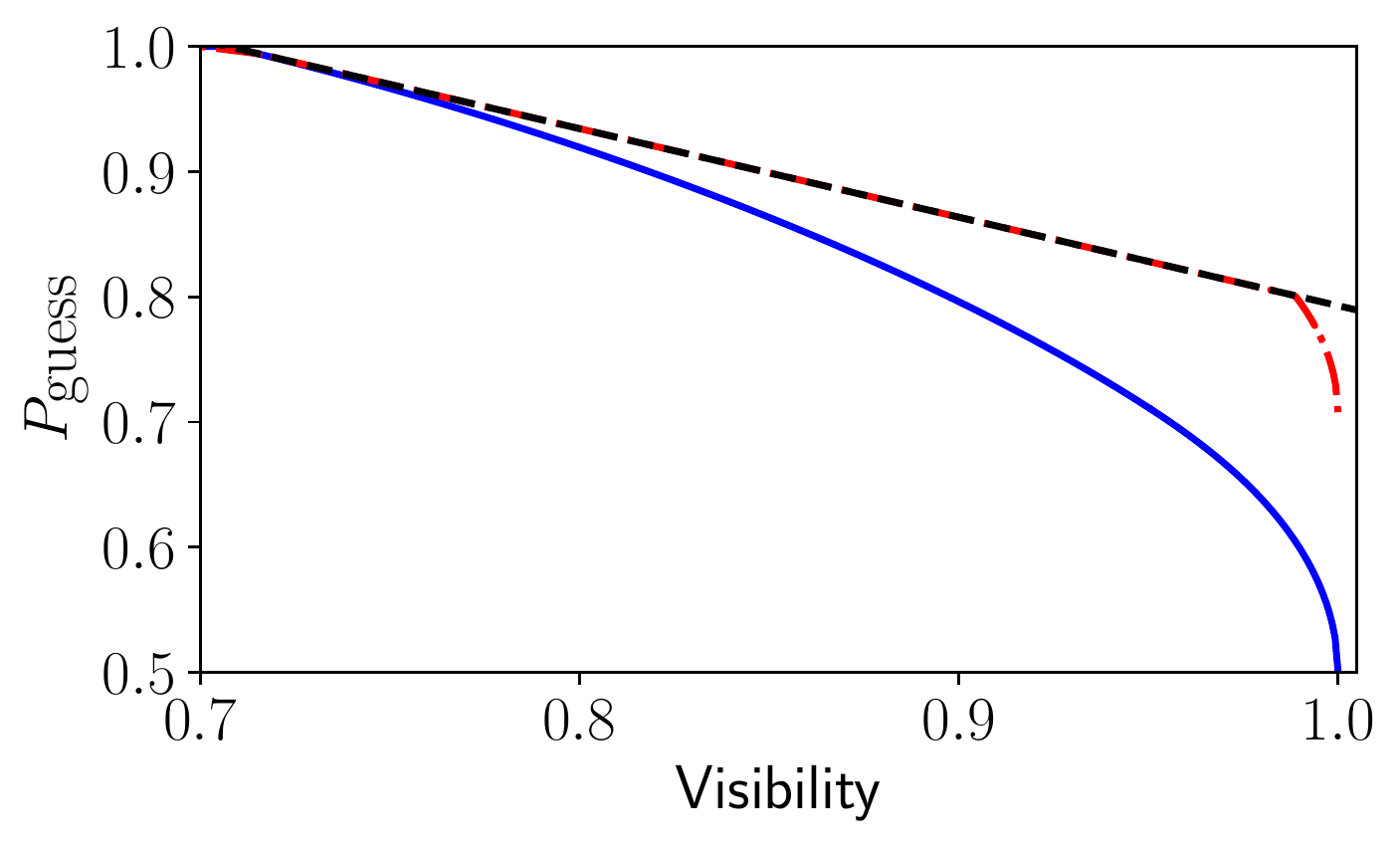}
    \caption{\textbf{Guessing probability as a function of the quantum state visibility.} We consider the parties share a state $\rho_v= v \ket{GHZ}\bra{GHZ} +(1-v)\openone/8$ and perform the optimal measurements maximizing the violation of \eqref{snew} given by $S(v)=4\sqrt{2}v$. The critical visibility value below which no violation is possible is given by $v_{\mathrm{crit}}=1/\sqrt{2}$ and as we see, above the critical value $p(e=c\vert x^*y^*z^*) <1$.  The straight blue curve gives the guessing probability $p(e=c\vert x^*y^*z^*)$ considering the level 2+ABC+ABE+BCE+ABCE in the semi-definite approximation of the quantum set \eqref{qconditions},  the dot-dashed curve in red shows the second level of approximation without these extra terms and the dashed straight line considers the NS set \eqref{NSconditions} (see the Appendix for more details).}
    \label{fig:plot2}
\end{figure}

The problem of security in the secret sharing protocol is thus equivalent to an optimization problem: find the maximum value for $p(e=c\vert x^*y^*z^*)$ (where $e$ stand for the extra outcome of Alice or Bob, depending on who is the untrusted part) given that a value of $S$ is observed and subject to the causal conditions of the model in Fig. \ref{fig:causals}c-d. As detailed in the Appendix and shown in Fig. \ref{fig:plot1}, the amount of information the untrusted part can get about the secret bit is a decreasing function of the violation of the Svetlichny inequality. Interestingly, however, the guessing probability only starts to decrease after the $S$ parameter surpass some threshold value ($S\approx 4.82$ for quantum and $S=6$ for NS correlations). Notwithstanding, below these threshold values, the security in the secret sharing protocol can be achived by looking at the full probability distribution instead of the single parameter $S$. This is illustrate in Fig. \ref{fig:plot2} by considering a GHZ state with a given visibility and measurements maximally violating the Svetlichny inequality \eqref{snew}. Further, as an example of a NS distribution fulfilling the secret sharing condition we have
\begin{equation}
p_v(a,b,c\vert x,y,z)= v p_{\mathrm{Svet}}+(1-v)p_{\mathrm{Clas}},
\end{equation}
with $p_{\mathrm{Svet}}(a,b,c\vert x,y,z)=(1/4)\delta_{a \oplus b \oplus c,xy \oplus xz \oplus yz}$ achieving the maximum Svetlichny violation $S=8$, $p_{\mathrm{Clas}}(a,b,c\vert x,y,z)=(1/4)\delta_{a \oplus b \oplus c,x}$ being classically correlated and achieving $S=4$. Thus $S(v)=4(1+v)$ and by solving the corresponding optimization problem one can show that $\max p(e=c\vert x^*y^*z^*)=1-v/2$, that is, in this case the guessing probability is a linearly decreasing function of the Svetlichny violation.

\emph{Discussion--} Quantum correlations allow for the ultimate level of security in communication protocols, since it is based on the own laws of physics and is achieved without the need of a precise knowledge of the physical apparatus being used. At the center of the device-independent approach for secure communication is the violation of a Bell inequality, however, the use as a resource of non-local correlations beyond the standard Bell definition is almost an uncharted territory. Here, we show the practical relevance of stronger than Bell non-local correlations, as they allow for the DI implementation of secret sharing, a communication primitive where the parties have to collaborate in order to recover a message. More precisely, attacks by an untrusted part can be modelled by causal models relaxing the measurement independence/locality assumption in Bell's theorem. Correlations incompatible with such nonlocal hidden variable models put strict bounds on the information achievable by the untrusted part and, in particular, the maximum violation of the Svetlichny inequality guarantees full security in the protocol.

An interesting venue for future research is to understand what other forms of non-classical behaviour can also be turned into resource for DI quantum information processing. As pointed out in \cite{chaves2017causal}, already at the tripartite scenario there are eight nonequivalent classes of non-local correlations, among which is the standard Bell definition and the stronger form we consider here. Can it be that for each class one can define a DI protocol? We believe the answer to that question can lead to even more exciting possibilities in quantum communication and we hope the present results may trigger more research in this direction.

\section*{Acknowledgements}
The authors acknowledge the Brazilian ministries MEC and MCTIC, funding agency CNPq (Grants No. 307172/2017-1 and No. 406574/2018-9 and INCT-IQ and PDJ Grant No. 154354/2018-0)). This work was supported by the Serrapilheira Institute (grant number Serra-1708-15763) and the John Templeton Foundation via the grant Q-CAUSAL No 61084.

\bibliography{Ref.bib}

\begin{thebibliography}{48}%
\makeatletter
\providecommand \@ifxundefined [1]{%
 \@ifx{#1\undefined}
}%
\providecommand \@ifnum [1]{%
 \ifnum #1\expandafter \@firstoftwo
 \else \expandafter \@secondoftwo
 \fi
}%
\providecommand \@ifx [1]{%
 \ifx #1\expandafter \@firstoftwo
 \else \expandafter \@secondoftwo
 \fi
}%
\providecommand \natexlab [1]{#1}%
\providecommand \enquote  [1]{``#1''}%
\providecommand \bibnamefont  [1]{#1}%
\providecommand \bibfnamefont [1]{#1}%
\providecommand \citenamefont [1]{#1}%
\providecommand \href@noop [0]{\@secondoftwo}%
\providecommand \href [0]{\begingroup \@sanitize@url \@href}%
\providecommand \@href[1]{\@@startlink{#1}\@@href}%
\providecommand \@@href[1]{\endgroup#1\@@endlink}%
\providecommand \@sanitize@url [0]{\catcode `\\12\catcode `\$12\catcode
  `\&12\catcode `\#12\catcode `\^12\catcode `\_12\catcode `\%12\relax}%
\providecommand \@@startlink[1]{}%
\providecommand \@@endlink[0]{}%
\providecommand \url  [0]{\begingroup\@sanitize@url \@url }%
\providecommand \@url [1]{\endgroup\@href {#1}{\urlprefix }}%
\providecommand \urlprefix  [0]{URL }%
\providecommand \Eprint [0]{\href }%
\providecommand \doibase [0]{http://dx.doi.org/}%
\providecommand \selectlanguage [0]{\@gobble}%
\providecommand \bibinfo  [0]{\@secondoftwo}%
\providecommand \bibfield  [0]{\@secondoftwo}%
\providecommand \translation [1]{[#1]}%
\providecommand \BibitemOpen [0]{}%
\providecommand \bibitemStop [0]{}%
\providecommand \bibitemNoStop [0]{.\EOS\space}%
\providecommand \EOS [0]{\spacefactor3000\relax}%
\providecommand \BibitemShut  [1]{\csname bibitem#1\endcsname}%
\let\auto@bib@innerbib\@empty
%</preamble>
\bibitem [{\citenamefont {Gisin}\ and\ \citenamefont
  {Thew}(2007)}]{gisin2007quantum}%
  \BibitemOpen
  \bibfield  {author} {\bibinfo {author} {\bibfnamefont {Nicolas}\ \bibnamefont
  {Gisin}}\ and\ \bibinfo {author} {\bibfnamefont {Rob}\ \bibnamefont {Thew}},\
  }\bibfield  {title} {\enquote {\bibinfo {title} {Quantum communication},}\
  }\href {\doibase 10.1038/nphoton.2007.22} {\bibfield  {journal} {\bibinfo
  {journal} {Nature Photonics}\ }\textbf {\bibinfo {volume} {1}},\ \bibinfo
  {pages} {165--171} (\bibinfo {year} {2007})}\BibitemShut {NoStop}%
\bibitem [{\citenamefont {Gisin}\ \emph {et~al.}(2002)\citenamefont {Gisin},
  \citenamefont {Ribordy}, \citenamefont {Tittel},\ and\ \citenamefont
  {Zbinden}}]{gisin2002quantum}%
  \BibitemOpen
  \bibfield  {author} {\bibinfo {author} {\bibfnamefont {Nicolas}\ \bibnamefont
  {Gisin}}, \bibinfo {author} {\bibfnamefont {Gr\'egoire}\ \bibnamefont
  {Ribordy}}, \bibinfo {author} {\bibfnamefont {Wolfgang}\ \bibnamefont
  {Tittel}}, \ and\ \bibinfo {author} {\bibfnamefont {Hugo}\ \bibnamefont
  {Zbinden}},\ }\bibfield  {title} {\enquote {\bibinfo {title} {Quantum
  cryptography},}\ }\href {\doibase 10.1103/RevModPhys.74.145} {\bibfield
  {journal} {\bibinfo  {journal} {Rev. Mod. Phys.}\ }\textbf {\bibinfo {volume}
  {74}},\ \bibinfo {pages} {145--195} (\bibinfo {year} {2002})}\BibitemShut
  {NoStop}%
\bibitem [{\citenamefont {Hensen}\ \emph {et~al.}(2015)\citenamefont {Hensen},
  \citenamefont {Bernien}, \citenamefont {Dr{\'e}au}, \citenamefont {Reiserer},
  \citenamefont {Kalb}, \citenamefont {Blok}, \citenamefont {Ruitenberg},
  \citenamefont {Vermeulen}, \citenamefont {Schouten}, \citenamefont
  {Abell{\'a}n}, \citenamefont {Amaya}, \citenamefont {Pruneri}, \citenamefont
  {Mitchell}, \citenamefont {Markham}, \citenamefont {Twitchen}, \citenamefont
  {Elkouss}, \citenamefont {Wehner}, \citenamefont {Taminiau},\ and\
  \citenamefont {Hanson}}]{Hensen2015}%
  \BibitemOpen
  \bibfield  {author} {\bibinfo {author} {\bibfnamefont {B.}~\bibnamefont
  {Hensen}}, \bibinfo {author} {\bibfnamefont {H.}~\bibnamefont {Bernien}},
  \bibinfo {author} {\bibfnamefont {A.~E.}\ \bibnamefont {Dr{\'e}au}}, \bibinfo
  {author} {\bibfnamefont {A.}~\bibnamefont {Reiserer}}, \bibinfo {author}
  {\bibfnamefont {N.}~\bibnamefont {Kalb}}, \bibinfo {author} {\bibfnamefont
  {M.~S.}\ \bibnamefont {Blok}}, \bibinfo {author} {\bibfnamefont
  {J.}~\bibnamefont {Ruitenberg}}, \bibinfo {author} {\bibfnamefont {R.~F.~L.}\
  \bibnamefont {Vermeulen}}, \bibinfo {author} {\bibfnamefont {R.~N.}\
  \bibnamefont {Schouten}}, \bibinfo {author} {\bibfnamefont {C.}~\bibnamefont
  {Abell{\'a}n}}, \bibinfo {author} {\bibfnamefont {W.}~\bibnamefont {Amaya}},
  \bibinfo {author} {\bibfnamefont {V.}~\bibnamefont {Pruneri}}, \bibinfo
  {author} {\bibfnamefont {M.~W.}\ \bibnamefont {Mitchell}}, \bibinfo {author}
  {\bibfnamefont {M.}~\bibnamefont {Markham}}, \bibinfo {author} {\bibfnamefont
  {D.~J.}\ \bibnamefont {Twitchen}}, \bibinfo {author} {\bibfnamefont
  {D.}~\bibnamefont {Elkouss}}, \bibinfo {author} {\bibfnamefont
  {S.}~\bibnamefont {Wehner}}, \bibinfo {author} {\bibfnamefont {T.~H.}\
  \bibnamefont {Taminiau}}, \ and\ \bibinfo {author} {\bibfnamefont
  {R.}~\bibnamefont {Hanson}},\ }\bibfield  {title} {\enquote {\bibinfo {title}
  {Loophole-free bell inequality violation using electron spins separated by
  1.3 kilometres},}\ }\href {https://doi.org/10.1038/nature15759} {\bibfield
  {journal} {\bibinfo  {journal} {Nature}\ }\textbf {\bibinfo {volume} {526}},\
  \bibinfo {pages} {682 EP --} (\bibinfo {year} {2015})}\BibitemShut {NoStop}%
\bibitem [{\citenamefont {Giustina}\ and\ \citenamefont {\emph{et
  al}}(2015)}]{Giustina2015}%
  \BibitemOpen
  \bibfield  {author} {\bibinfo {author} {\bibfnamefont {M.}~\bibnamefont
  {Giustina}}\ and\ \bibinfo {author} {\bibnamefont {\emph{et al}}},\
  }\bibfield  {title} {\enquote {\bibinfo {title} {Significant-loophole-free
  test of bell's theorem with entangled photons},}\ }\href {\doibase
  10.1103/PhysRevLett.115.250401} {\bibfield  {journal} {\bibinfo  {journal}
  {Phys. Rev. Lett.}\ }\textbf {\bibinfo {volume} {115}},\ \bibinfo {pages}
  {250401} (\bibinfo {year} {2015})}\BibitemShut {NoStop}%
\bibitem [{\citenamefont {Shalm}\ and\ \citenamefont {\emph{et
  al}}(2015)}]{Shalm2015}%
  \BibitemOpen
  \bibfield  {author} {\bibinfo {author} {\bibfnamefont {L.~K.}\ \bibnamefont
  {Shalm}}\ and\ \bibinfo {author} {\bibnamefont {\emph{et al}}},\ }\bibfield
  {title} {\enquote {\bibinfo {title} {Strong loophole-free test of local
  realism},}\ }\href {\doibase 10.1103/PhysRevLett.115.250402} {\bibfield
  {journal} {\bibinfo  {journal} {Phys. Rev. Lett.}\ }\textbf {\bibinfo
  {volume} {115}},\ \bibinfo {pages} {250402} (\bibinfo {year}
  {2015})}\BibitemShut {NoStop}%
\bibitem [{\citenamefont {Yin}\ \emph {et~al.}(2017)\citenamefont {Yin},
  \citenamefont {Cao}, \citenamefont {Li}, \citenamefont {Liao}, \citenamefont
  {Zhang}, \citenamefont {Ren}, \citenamefont {Cai}, \citenamefont {Liu},
  \citenamefont {Li}, \citenamefont {Dai}, \citenamefont {Li}, \citenamefont
  {Lu}, \citenamefont {Gong}, \citenamefont {Xu}, \citenamefont {Li},
  \citenamefont {Li}, \citenamefont {Yin}, \citenamefont {Jiang}, \citenamefont
  {Li}, \citenamefont {Jia}, \citenamefont {Ren}, \citenamefont {He},
  \citenamefont {Zhou}, \citenamefont {Zhang}, \citenamefont {Wang},
  \citenamefont {Chang}, \citenamefont {Zhu}, \citenamefont {Liu},
  \citenamefont {Chen}, \citenamefont {Lu}, \citenamefont {Shu}, \citenamefont
  {Peng}, \citenamefont {Wang},\ and\ \citenamefont {Pan}}]{Yin2017}%
  \BibitemOpen
  \bibfield  {author} {\bibinfo {author} {\bibfnamefont {Juan}\ \bibnamefont
  {Yin}}, \bibinfo {author} {\bibfnamefont {Yuan}\ \bibnamefont {Cao}},
  \bibinfo {author} {\bibfnamefont {Yu-Huai}\ \bibnamefont {Li}}, \bibinfo
  {author} {\bibfnamefont {Sheng-Kai}\ \bibnamefont {Liao}}, \bibinfo {author}
  {\bibfnamefont {Liang}\ \bibnamefont {Zhang}}, \bibinfo {author}
  {\bibfnamefont {Ji-Gang}\ \bibnamefont {Ren}}, \bibinfo {author}
  {\bibfnamefont {Wen-Qi}\ \bibnamefont {Cai}}, \bibinfo {author}
  {\bibfnamefont {Wei-Yue}\ \bibnamefont {Liu}}, \bibinfo {author}
  {\bibfnamefont {Bo}~\bibnamefont {Li}}, \bibinfo {author} {\bibfnamefont
  {Hui}\ \bibnamefont {Dai}}, \bibinfo {author} {\bibfnamefont {Guang-Bing}\
  \bibnamefont {Li}}, \bibinfo {author} {\bibfnamefont {Qi-Ming}\ \bibnamefont
  {Lu}}, \bibinfo {author} {\bibfnamefont {Yun-Hong}\ \bibnamefont {Gong}},
  \bibinfo {author} {\bibfnamefont {Yu}~\bibnamefont {Xu}}, \bibinfo {author}
  {\bibfnamefont {Shuang-Lin}\ \bibnamefont {Li}}, \bibinfo {author}
  {\bibfnamefont {Feng-Zhi}\ \bibnamefont {Li}}, \bibinfo {author}
  {\bibfnamefont {Ya-Yun}\ \bibnamefont {Yin}}, \bibinfo {author}
  {\bibfnamefont {Zi-Qing}\ \bibnamefont {Jiang}}, \bibinfo {author}
  {\bibfnamefont {Ming}\ \bibnamefont {Li}}, \bibinfo {author} {\bibfnamefont
  {Jian-Jun}\ \bibnamefont {Jia}}, \bibinfo {author} {\bibfnamefont
  {Ge}~\bibnamefont {Ren}}, \bibinfo {author} {\bibfnamefont {Dong}\
  \bibnamefont {He}}, \bibinfo {author} {\bibfnamefont {Yi-Lin}\ \bibnamefont
  {Zhou}}, \bibinfo {author} {\bibfnamefont {Xiao-Xiang}\ \bibnamefont
  {Zhang}}, \bibinfo {author} {\bibfnamefont {Na}~\bibnamefont {Wang}},
  \bibinfo {author} {\bibfnamefont {Xiang}\ \bibnamefont {Chang}}, \bibinfo
  {author} {\bibfnamefont {Zhen-Cai}\ \bibnamefont {Zhu}}, \bibinfo {author}
  {\bibfnamefont {Nai-Le}\ \bibnamefont {Liu}}, \bibinfo {author}
  {\bibfnamefont {Yu-Ao}\ \bibnamefont {Chen}}, \bibinfo {author}
  {\bibfnamefont {Chao-Yang}\ \bibnamefont {Lu}}, \bibinfo {author}
  {\bibfnamefont {Rong}\ \bibnamefont {Shu}}, \bibinfo {author} {\bibfnamefont
  {Cheng-Zhi}\ \bibnamefont {Peng}}, \bibinfo {author} {\bibfnamefont
  {Jian-Yu}\ \bibnamefont {Wang}}, \ and\ \bibinfo {author} {\bibfnamefont
  {Jian-Wei}\ \bibnamefont {Pan}},\ }\bibfield  {title} {\enquote {\bibinfo
  {title} {Satellite-based entanglement distribution over 1200 kilometers},}\
  }\href {\doibase 10.1126/science.aan3211} {\bibfield  {journal} {\bibinfo
  {journal} {Science}\ }\textbf {\bibinfo {volume} {356}},\ \bibinfo {pages}
  {1140--1144} (\bibinfo {year} {2017})}\BibitemShut {NoStop}%
\bibitem [{\citenamefont {Kimble}(2008)}]{Kimble2008}%
  \BibitemOpen
  \bibfield  {author} {\bibinfo {author} {\bibfnamefont {H.~J.}\ \bibnamefont
  {Kimble}},\ }\bibfield  {title} {\enquote {\bibinfo {title} {The quantum
  internet},}\ }\href {\doibase 10.1038/nature07127} {\bibfield  {journal}
  {\bibinfo  {journal} {Nature}\ }\textbf {\bibinfo {volume} {453}},\ \bibinfo
  {pages} {1023--1030} (\bibinfo {year} {2008})}\BibitemShut {NoStop}%
\bibitem [{\citenamefont {Ekert}(1991)}]{Ekert1991}%
  \BibitemOpen
  \bibfield  {author} {\bibinfo {author} {\bibfnamefont {Artur~K.}\
  \bibnamefont {Ekert}},\ }\bibfield  {title} {\enquote {\bibinfo {title}
  {Quantum cryptography based on bell's theorem},}\ }\href {\doibase
  10.1103/PhysRevLett.67.661} {\bibfield  {journal} {\bibinfo  {journal} {Phys.
  Rev. Lett.}\ }\textbf {\bibinfo {volume} {67}},\ \bibinfo {pages} {661--663}
  (\bibinfo {year} {1991})}\BibitemShut {NoStop}%
\bibitem [{\citenamefont {Bennett}\ and\ \citenamefont
  {Brassard}(2014)}]{bennett2014quantum}%
  \BibitemOpen
  \bibfield  {author} {\bibinfo {author} {\bibfnamefont {Charles~H.}\
  \bibnamefont {Bennett}}\ and\ \bibinfo {author} {\bibfnamefont {Gilles}\
  \bibnamefont {Brassard}},\ }\bibfield  {title} {\enquote {\bibinfo {title}
  {Quantum cryptography: Public key distribution and coin tossing},}\ }\href
  {\doibase https://doi.org/10.1016/j.tcs.2014.05.025} {\bibfield  {journal}
  {\bibinfo  {journal} {Theoretical Computer Science}\ }\textbf {\bibinfo
  {volume} {560}},\ \bibinfo {pages} {7 -- 11} (\bibinfo {year} {2014})},\
  \bibinfo {note} {theoretical Aspects of Quantum Cryptography – celebrating
  30 years of BB84}\BibitemShut {NoStop}%
\bibitem [{\citenamefont {Lydersen}\ \emph {et~al.}(2010)\citenamefont
  {Lydersen}, \citenamefont {Wiechers}, \citenamefont {Wittmann}, \citenamefont
  {Elser}, \citenamefont {Skaar},\ and\ \citenamefont
  {Makarov}}]{lydersen2010hacking}%
  \BibitemOpen
  \bibfield  {author} {\bibinfo {author} {\bibfnamefont {Lars}\ \bibnamefont
  {Lydersen}}, \bibinfo {author} {\bibfnamefont {Carlos}\ \bibnamefont
  {Wiechers}}, \bibinfo {author} {\bibfnamefont {Christoffer}\ \bibnamefont
  {Wittmann}}, \bibinfo {author} {\bibfnamefont {Dominique}\ \bibnamefont
  {Elser}}, \bibinfo {author} {\bibfnamefont {Johannes}\ \bibnamefont {Skaar}},
  \ and\ \bibinfo {author} {\bibfnamefont {Vadim}\ \bibnamefont {Makarov}},\
  }\bibfield  {title} {\enquote {\bibinfo {title} {Hacking commercial quantum
  cryptography systems by tailored bright illumination},}\ }\href {\doibase
  10.1038/nphoton.2010.214} {\bibfield  {journal} {\bibinfo  {journal} {Nature
  Photonics}\ }\textbf {\bibinfo {volume} {4}},\ \bibinfo {pages} {686--689}
  (\bibinfo {year} {2010})}\BibitemShut {NoStop}%
\bibitem [{\citenamefont {{Mayers}}\ and\ \citenamefont
  {{Yao}}(1998)}]{mayers1998quantum}%
  \BibitemOpen
  \bibfield  {author} {\bibinfo {author} {\bibfnamefont {D.}~\bibnamefont
  {{Mayers}}}\ and\ \bibinfo {author} {\bibfnamefont {A.}~\bibnamefont
  {{Yao}}},\ }\bibfield  {title} {\enquote {\bibinfo {title} {Quantum
  cryptography with imperfect apparatus},}\ }\bibfield  {booktitle} {\emph
  {\bibinfo {booktitle} {Proceedings 39th Annual Symposium on Foundations of
  Computer Science (Cat. No.98CB36280)}},\ }\href {\doibase
  10.1109/SFCS.1998.743501} {\ ,\ \bibinfo {pages} {503--509} (\bibinfo {year}
  {1998})}\BibitemShut {NoStop}%
\bibitem [{\citenamefont {Barrett}\ \emph {et~al.}(2005)\citenamefont
  {Barrett}, \citenamefont {Hardy},\ and\ \citenamefont
  {Kent}}]{PhysRevLett.95.010503}%
  \BibitemOpen
  \bibfield  {author} {\bibinfo {author} {\bibfnamefont {Jonathan}\
  \bibnamefont {Barrett}}, \bibinfo {author} {\bibfnamefont {Lucien}\
  \bibnamefont {Hardy}}, \ and\ \bibinfo {author} {\bibfnamefont {Adrian}\
  \bibnamefont {Kent}},\ }\bibfield  {title} {\enquote {\bibinfo {title} {No
  signaling and quantum key distribution},}\ }\href {\doibase
  10.1103/PhysRevLett.95.010503} {\bibfield  {journal} {\bibinfo  {journal}
  {Phys. Rev. Lett.}\ }\textbf {\bibinfo {volume} {95}},\ \bibinfo {pages}
  {010503} (\bibinfo {year} {2005})}\BibitemShut {NoStop}%
\bibitem [{\citenamefont {Vazirani}\ and\ \citenamefont
  {Vidick}(2014)}]{PhysRevLett.113.140501}%
  \BibitemOpen
  \bibfield  {author} {\bibinfo {author} {\bibfnamefont {Umesh}\ \bibnamefont
  {Vazirani}}\ and\ \bibinfo {author} {\bibfnamefont {Thomas}\ \bibnamefont
  {Vidick}},\ }\bibfield  {title} {\enquote {\bibinfo {title} {Fully
  device-independent quantum key distribution},}\ }\href {\doibase
  10.1103/PhysRevLett.113.140501} {\bibfield  {journal} {\bibinfo  {journal}
  {Phys. Rev. Lett.}\ }\textbf {\bibinfo {volume} {113}},\ \bibinfo {pages}
  {140501} (\bibinfo {year} {2014})}\BibitemShut {NoStop}%
\bibitem [{\citenamefont {Bell}(1964)}]{Bell1964}%
  \BibitemOpen
  \bibfield  {author} {\bibinfo {author} {\bibfnamefont {J.~S.}\ \bibnamefont
  {Bell}},\ }\bibfield  {title} {\enquote {\bibinfo {title} {On the einstein
  podolsky rosen paradox},}\ }\href {\doibase
  10.1103/PhysicsPhysiqueFizika.1.195} {\bibfield  {journal} {\bibinfo
  {journal} {Physics Physique Fizika}\ }\textbf {\bibinfo {volume} {1}},\
  \bibinfo {pages} {195--200} (\bibinfo {year} {1964})}\BibitemShut {NoStop}%
\bibitem [{\citenamefont {Colbeck}(2007)}]{Colbeck2007}%
  \BibitemOpen
  \bibfield  {author} {\bibinfo {author} {\bibfnamefont {R.}~\bibnamefont
  {Colbeck}},\ }\emph {\bibinfo {title} {Ph.D.~Thesis}},\ \href@noop {} {Ph.D.
  thesis},\ \bibinfo  {school} {University of Cambridge} (\bibinfo {year}
  {2007})\BibitemShut {NoStop}%
\bibitem [{\citenamefont {Pironio}\ \emph {et~al.}(2010)\citenamefont
  {Pironio}, \citenamefont {Ac{\'\i}n}, \citenamefont {Massar}, \citenamefont
  {de~la Giroday}, \citenamefont {Matsukevich}, \citenamefont {Maunz},
  \citenamefont {Olmschenk}, \citenamefont {Hayes}, \citenamefont {Luo},
  \citenamefont {Manning},\ and\ \citenamefont {Monroe}}]{Pironio2010}%
  \BibitemOpen
  \bibfield  {author} {\bibinfo {author} {\bibfnamefont {S.}~\bibnamefont
  {Pironio}}, \bibinfo {author} {\bibfnamefont {A.}~\bibnamefont {Ac{\'\i}n}},
  \bibinfo {author} {\bibfnamefont {S.}~\bibnamefont {Massar}}, \bibinfo
  {author} {\bibfnamefont {A.~Boyer}\ \bibnamefont {de~la Giroday}}, \bibinfo
  {author} {\bibfnamefont {D.~N.}\ \bibnamefont {Matsukevich}}, \bibinfo
  {author} {\bibfnamefont {P.}~\bibnamefont {Maunz}}, \bibinfo {author}
  {\bibfnamefont {S.}~\bibnamefont {Olmschenk}}, \bibinfo {author}
  {\bibfnamefont {D.}~\bibnamefont {Hayes}}, \bibinfo {author} {\bibfnamefont
  {L.}~\bibnamefont {Luo}}, \bibinfo {author} {\bibfnamefont {T.~A.}\
  \bibnamefont {Manning}}, \ and\ \bibinfo {author} {\bibfnamefont
  {C.}~\bibnamefont {Monroe}},\ }\bibfield  {title} {\enquote {\bibinfo {title}
  {Random numbers certified by bell's theorem},}\ }\href
  {https://doi.org/10.1038/nature09008} {\bibfield  {journal} {\bibinfo
  {journal} {Nature}\ }\textbf {\bibinfo {volume} {464}},\ \bibinfo {pages}
  {1021 EP --} (\bibinfo {year} {2010})}\BibitemShut {NoStop}%
\bibitem [{\citenamefont {Pearl}(2009)}]{Pearl2009}%
  \BibitemOpen
  \bibfield  {author} {\bibinfo {author} {\bibfnamefont {J.}~\bibnamefont
  {Pearl}},\ }\href@noop {} {\emph {\bibinfo {title} {Causality}}}\ (\bibinfo
  {publisher} {Cambridge University Press},\ \bibinfo {year}
  {2009})\BibitemShut {NoStop}%
\bibitem [{\citenamefont {Fritz}(2016)}]{Fritz2014}%
  \BibitemOpen
  \bibfield  {author} {\bibinfo {author} {\bibfnamefont {T.}~\bibnamefont
  {Fritz}},\ }\bibfield  {title} {\enquote {\bibinfo {title} {Beyond bell's
  theorem ii: Scenarios with arbitrary causal structure},}\ }\href {\doibase
  10.1007/s00220-015-2495-5} {\bibfield  {journal} {\bibinfo  {journal}
  {Communications in Mathematical Physics}\ }\textbf {\bibinfo {volume}
  {341}},\ \bibinfo {pages} {391--434} (\bibinfo {year} {2016})}\BibitemShut
  {NoStop}%
\bibitem [{\citenamefont {Wood}\ and\ \citenamefont
  {Spekkens}(2015)}]{Spekkens2015}%
  \BibitemOpen
  \bibfield  {author} {\bibinfo {author} {\bibfnamefont {C.~J.}\ \bibnamefont
  {Wood}}\ and\ \bibinfo {author} {\bibfnamefont {R.~W.}\ \bibnamefont
  {Spekkens}},\ }\bibfield  {title} {\enquote {\bibinfo {title} {The lesson of
  causal discovery algorithms for quantum correlations: causal explanations of
  {B}ell-inequality violations require fine-tuning},}\ }\href
  {http://stacks.iop.org/1367-2630/17/i=3/a=033002} {\bibfield  {journal}
  {\bibinfo  {journal} {New J. Phys.}\ }\textbf {\bibinfo {volume} {17}},\
  \bibinfo {pages} {033002} (\bibinfo {year} {2015})}\BibitemShut {NoStop}%
\bibitem [{\citenamefont {Chaves}\ \emph {et~al.}(2015)\citenamefont {Chaves},
  \citenamefont {Kueng}, \citenamefont {Brask},\ and\ \citenamefont
  {Gross}}]{Chaves2015b}%
  \BibitemOpen
  \bibfield  {author} {\bibinfo {author} {\bibfnamefont {R.}~\bibnamefont
  {Chaves}}, \bibinfo {author} {\bibfnamefont {R.}~\bibnamefont {Kueng}},
  \bibinfo {author} {\bibfnamefont {J.~B.}\ \bibnamefont {Brask}}, \ and\
  \bibinfo {author} {\bibfnamefont {D.}~\bibnamefont {Gross}},\ }\bibfield
  {title} {\enquote {\bibinfo {title} {Unifying framework for relaxations of
  the causal assumptions in {B}ell's theorem},}\ }\href {\doibase
  10.1103/PhysRevLett.114.140403} {\bibfield  {journal} {\bibinfo  {journal}
  {Phys. Rev. Lett.}\ }\textbf {\bibinfo {volume} {114}},\ \bibinfo {pages}
  {140403} (\bibinfo {year} {2015})}\BibitemShut {NoStop}%
\bibitem [{\citenamefont {Renou}\ \emph {et~al.}(2019)\citenamefont {Renou},
  \citenamefont {B{\"a}umer}, \citenamefont {Boreiri}, \citenamefont {Brunner},
  \citenamefont {Gisin},\ and\ \citenamefont {Beigi}}]{renou2019genuine}%
  \BibitemOpen
  \bibfield  {author} {\bibinfo {author} {\bibfnamefont {Marc-Olivier}\
  \bibnamefont {Renou}}, \bibinfo {author} {\bibfnamefont {Elisa}\ \bibnamefont
  {B{\"a}umer}}, \bibinfo {author} {\bibfnamefont {Sadra}\ \bibnamefont
  {Boreiri}}, \bibinfo {author} {\bibfnamefont {Nicolas}\ \bibnamefont
  {Brunner}}, \bibinfo {author} {\bibfnamefont {Nicolas}\ \bibnamefont
  {Gisin}}, \ and\ \bibinfo {author} {\bibfnamefont {Salman}\ \bibnamefont
  {Beigi}},\ }\bibfield  {title} {\enquote {\bibinfo {title} {Genuine quantum
  nonlocality in the triangle network},}\ }\href@noop {} {\bibfield  {journal}
  {\bibinfo  {journal} {arXiv preprint arXiv:1905.04902}\ } (\bibinfo {year}
  {2019})}\BibitemShut {NoStop}%
\bibitem [{\citenamefont {Brunner}\ \emph {et~al.}(2014)\citenamefont
  {Brunner}, \citenamefont {Cavalcanti}, \citenamefont {Pironio}, \citenamefont
  {Scarani},\ and\ \citenamefont {Wehner}}]{Brunner2014}%
  \BibitemOpen
  \bibfield  {author} {\bibinfo {author} {\bibfnamefont {N.}~\bibnamefont
  {Brunner}}, \bibinfo {author} {\bibfnamefont {D.}~\bibnamefont {Cavalcanti}},
  \bibinfo {author} {\bibfnamefont {S.}~\bibnamefont {Pironio}}, \bibinfo
  {author} {\bibfnamefont {V.}~\bibnamefont {Scarani}}, \ and\ \bibinfo
  {author} {\bibfnamefont {S.}~\bibnamefont {Wehner}},\ }\bibfield  {title}
  {\enquote {\bibinfo {title} {Bell nonlocality},}\ }\href {\doibase
  10.1103/RevModPhys.86.419} {\bibfield  {journal} {\bibinfo  {journal} {Rev.
  Mod. Phys.}\ }\textbf {\bibinfo {volume} {86}},\ \bibinfo {pages} {419--478}
  (\bibinfo {year} {2014})}\BibitemShut {NoStop}%
\bibitem [{\citenamefont {Lee}\ and\ \citenamefont
  {Hoban}(2018)}]{lee2018towards}%
  \BibitemOpen
  \bibfield  {author} {\bibinfo {author} {\bibfnamefont {Ciar\'an~M.}\
  \bibnamefont {Lee}}\ and\ \bibinfo {author} {\bibfnamefont {Matty~J.}\
  \bibnamefont {Hoban}},\ }\bibfield  {title} {\enquote {\bibinfo {title}
  {Towards device-independent information processing on general quantum
  networks},}\ }\href {\doibase 10.1103/PhysRevLett.120.020504} {\bibfield
  {journal} {\bibinfo  {journal} {Phys. Rev. Lett.}\ }\textbf {\bibinfo
  {volume} {120}},\ \bibinfo {pages} {020504} (\bibinfo {year}
  {2018})}\BibitemShut {NoStop}%
\bibitem [{\citenamefont {Chaves}\ \emph {et~al.}(2018)\citenamefont {Chaves},
  \citenamefont {Carvacho}, \citenamefont {Agresti}, \citenamefont {Di~Giulio},
  \citenamefont {Aolita}, \citenamefont {Giacomini},\ and\ \citenamefont
  {Sciarrino}}]{Chaves2018}%
  \BibitemOpen
  \bibfield  {author} {\bibinfo {author} {\bibfnamefont {Rafael}\ \bibnamefont
  {Chaves}}, \bibinfo {author} {\bibfnamefont {Gonzalo}\ \bibnamefont
  {Carvacho}}, \bibinfo {author} {\bibfnamefont {Iris}\ \bibnamefont
  {Agresti}}, \bibinfo {author} {\bibfnamefont {Valerio}\ \bibnamefont
  {Di~Giulio}}, \bibinfo {author} {\bibfnamefont {Leandro}\ \bibnamefont
  {Aolita}}, \bibinfo {author} {\bibfnamefont {Sandro}\ \bibnamefont
  {Giacomini}}, \ and\ \bibinfo {author} {\bibfnamefont {Fabio}\ \bibnamefont
  {Sciarrino}},\ }\bibfield  {title} {\enquote {\bibinfo {title} {Quantum
  violation of an instrumental test},}\ }\href {\doibase
  10.1038/s41567-017-0008-5} {\bibfield  {journal} {\bibinfo  {journal} {Nature
  Physics}\ }\textbf {\bibinfo {volume} {14}},\ \bibinfo {pages} {291--296}
  (\bibinfo {year} {2018})}\BibitemShut {NoStop}%
\bibitem [{\citenamefont {Agresti}\ \emph {et~al.}(2019)\citenamefont
  {Agresti}, \citenamefont {Poderini}, \citenamefont {Guerini}, \citenamefont
  {Mancusi}, \citenamefont {Carvacho}, \citenamefont {Aolita}, \citenamefont
  {Cavalcanti}, \citenamefont {Chaves},\ and\ \citenamefont
  {Sciarrino}}]{agresti2019experimental}%
  \BibitemOpen
  \bibfield  {author} {\bibinfo {author} {\bibfnamefont {Iris}\ \bibnamefont
  {Agresti}}, \bibinfo {author} {\bibfnamefont {Davide}\ \bibnamefont
  {Poderini}}, \bibinfo {author} {\bibfnamefont {Leonardo}\ \bibnamefont
  {Guerini}}, \bibinfo {author} {\bibfnamefont {Michele}\ \bibnamefont
  {Mancusi}}, \bibinfo {author} {\bibfnamefont {Gonzalo}\ \bibnamefont
  {Carvacho}}, \bibinfo {author} {\bibfnamefont {Leandro}\ \bibnamefont
  {Aolita}}, \bibinfo {author} {\bibfnamefont {Daniel}\ \bibnamefont
  {Cavalcanti}}, \bibinfo {author} {\bibfnamefont {Rafael}\ \bibnamefont
  {Chaves}}, \ and\ \bibinfo {author} {\bibfnamefont {Fabio}\ \bibnamefont
  {Sciarrino}},\ }\bibfield  {title} {\enquote {\bibinfo {title} {Experimental
  device-independent certified randomness generation with an instrumental
  causal structure},}\ }\href {https://arxiv.org/pdf/1905.02027} {\bibfield
  {journal} {\bibinfo  {journal} {arXiv preprint arXiv:1905.02027}\ } (\bibinfo
  {year} {2019})}\BibitemShut {NoStop}%
\bibitem [{\citenamefont {Hillery}\ \emph {et~al.}(1999)\citenamefont
  {Hillery}, \citenamefont {Buzek},\ and\ \citenamefont
  {Berthiaume}}]{PhysRevA.59.1829}%
  \BibitemOpen
  \bibfield  {author} {\bibinfo {author} {\bibfnamefont {Mark}\ \bibnamefont
  {Hillery}}, \bibinfo {author} {\bibfnamefont {Vladim{\'\i}r}\ \bibnamefont
  {Buzek}}, \ and\ \bibinfo {author} {\bibfnamefont {Andr{\'e}}\ \bibnamefont
  {Berthiaume}},\ }\bibfield  {title} {\enquote {\bibinfo {title} {Quantum
  secret sharing},}\ }\href {\doibase 10.1103/PhysRevA.59.1829} {\bibfield
  {journal} {\bibinfo  {journal} {Phys. Rev. A}\ }\textbf {\bibinfo {volume}
  {59}},\ \bibinfo {pages} {1829--1834} (\bibinfo {year} {1999})}\BibitemShut
  {NoStop}%
\bibitem [{\citenamefont {Karlsson}\ \emph {et~al.}(1999)\citenamefont
  {Karlsson}, \citenamefont {Koashi},\ and\ \citenamefont
  {Imoto}}]{PhysRevA.59.162}%
  \BibitemOpen
  \bibfield  {author} {\bibinfo {author} {\bibfnamefont {Anders}\ \bibnamefont
  {Karlsson}}, \bibinfo {author} {\bibfnamefont {Masato}\ \bibnamefont
  {Koashi}}, \ and\ \bibinfo {author} {\bibfnamefont {Nobuyuki}\ \bibnamefont
  {Imoto}},\ }\bibfield  {title} {\enquote {\bibinfo {title} {Quantum
  entanglement for secret sharing and secret splitting},}\ }\href {\doibase
  10.1103/PhysRevA.59.162} {\bibfield  {journal} {\bibinfo  {journal} {Phys.
  Rev. A}\ }\textbf {\bibinfo {volume} {59}},\ \bibinfo {pages} {162--168}
  (\bibinfo {year} {1999})}\BibitemShut {NoStop}%
\bibitem [{\citenamefont {Aolita}\ \emph {et~al.}(2012)\citenamefont {Aolita},
  \citenamefont {Gallego}, \citenamefont {Cabello},\ and\ \citenamefont
  {Ac\'{\i}n}}]{PhysRevLett.108.100401}%
  \BibitemOpen
  \bibfield  {author} {\bibinfo {author} {\bibfnamefont {Leandro}\ \bibnamefont
  {Aolita}}, \bibinfo {author} {\bibfnamefont {Rodrigo}\ \bibnamefont
  {Gallego}}, \bibinfo {author} {\bibfnamefont {Ad\'an}\ \bibnamefont
  {Cabello}}, \ and\ \bibinfo {author} {\bibfnamefont {Antonio}\ \bibnamefont
  {Ac\'{\i}n}},\ }\bibfield  {title} {\enquote {\bibinfo {title} {Fully
  nonlocal, monogamous, and random genuinely multipartite quantum
  correlations},}\ }\href {\doibase 10.1103/PhysRevLett.108.100401} {\bibfield
  {journal} {\bibinfo  {journal} {Phys. Rev. Lett.}\ }\textbf {\bibinfo
  {volume} {108}},\ \bibinfo {pages} {100401} (\bibinfo {year}
  {2012})}\BibitemShut {NoStop}%
\bibitem [{\citenamefont {Woodhead}\ \emph {et~al.}(2018)\citenamefont
  {Woodhead}, \citenamefont {Bourdoncle},\ and\ \citenamefont
  {Ac{\'\i}n}}]{woodhead2018randomness}%
  \BibitemOpen
  \bibfield  {author} {\bibinfo {author} {\bibfnamefont {Erik}\ \bibnamefont
  {Woodhead}}, \bibinfo {author} {\bibfnamefont {Boris}\ \bibnamefont
  {Bourdoncle}}, \ and\ \bibinfo {author} {\bibfnamefont {Antonio}\
  \bibnamefont {Ac{\'\i}n}},\ }\bibfield  {title} {\enquote {\bibinfo {title}
  {Randomness versus nonlocality in the mermin-bell experiment with three
  parties},}\ }\href@noop {} {\bibfield  {journal} {\bibinfo  {journal} {arXiv
  preprint arXiv:1804.09733}\ } (\bibinfo {year} {2018})}\BibitemShut {NoStop}%
\bibitem [{\citenamefont {Chaves}\ \emph {et~al.}(2017)\citenamefont {Chaves},
  \citenamefont {Cavalcanti},\ and\ \citenamefont {Aolita}}]{chaves2017causal}%
  \BibitemOpen
  \bibfield  {author} {\bibinfo {author} {\bibfnamefont {Rafael}\ \bibnamefont
  {Chaves}}, \bibinfo {author} {\bibfnamefont {Daniel}\ \bibnamefont
  {Cavalcanti}}, \ and\ \bibinfo {author} {\bibfnamefont {Leandro}\
  \bibnamefont {Aolita}},\ }\bibfield  {title} {\enquote {\bibinfo {title}
  {Causal hierarchy of multipartite {B}ell nonlocality},}\ }\href {\doibase
  10.22331/q-2017-08-04-23} {\bibfield  {journal} {\bibinfo  {journal}
  {{Quantum}}\ }\textbf {\bibinfo {volume} {1}},\ \bibinfo {pages} {23}
  (\bibinfo {year} {2017})}\BibitemShut {NoStop}%
\bibitem [{\citenamefont {Svetlichny}(1987)}]{Svet1987}%
  \BibitemOpen
  \bibfield  {author} {\bibinfo {author} {\bibfnamefont {George}\ \bibnamefont
  {Svetlichny}},\ }\bibfield  {title} {\enquote {\bibinfo {title}
  {Distinguishing three-body from two-body nonseparability by a bell-type
  inequality},}\ }\href {\doibase 10.1103/PhysRevD.35.3066} {\bibfield
  {journal} {\bibinfo  {journal} {Phys. Rev. D}\ }\textbf {\bibinfo {volume}
  {35}},\ \bibinfo {pages} {3066--3069} (\bibinfo {year} {1987})}\BibitemShut
  {NoStop}%
\bibitem [{\citenamefont {Popescu}\ and\ \citenamefont
  {Rohrlich}(1994)}]{Popescu1994}%
  \BibitemOpen
  \bibfield  {author} {\bibinfo {author} {\bibfnamefont {S.}~\bibnamefont
  {Popescu}}\ and\ \bibinfo {author} {\bibfnamefont {D.}~\bibnamefont
  {Rohrlich}},\ }\bibfield  {title} {\enquote {\bibinfo {title} {Quantum
  nonlocality as an axiom},}\ }\href {http://dx.doi.org/10.1007/BF02058098}
  {\bibfield  {journal} {\bibinfo  {journal} {Foundations of Physics}\ }\textbf
  {\bibinfo {volume} {24}},\ \bibinfo {pages} {379--385} (\bibinfo {year}
  {1994})}\BibitemShut {NoStop}%
\bibitem [{\citenamefont {Greenberger}\ \emph {et~al.}(1989)\citenamefont
  {Greenberger}, \citenamefont {Horne},\ and\ \citenamefont
  {Zeilinger}}]{greenberger1989going}%
  \BibitemOpen
  \bibfield  {author} {\bibinfo {author} {\bibfnamefont {Daniel~M}\
  \bibnamefont {Greenberger}}, \bibinfo {author} {\bibfnamefont {Michael~A}\
  \bibnamefont {Horne}}, \ and\ \bibinfo {author} {\bibfnamefont {Anton}\
  \bibnamefont {Zeilinger}},\ }\bibfield  {title} {\enquote {\bibinfo {title}
  {Going beyond bell’s theorem},}\ }in\ \href@noop {} {\emph {\bibinfo
  {booktitle} {Bell’s theorem, quantum theory and conceptions of the
  universe}}}\ (\bibinfo  {publisher} {Springer},\ \bibinfo {year} {1989})\
  pp.\ \bibinfo {pages} {69--72}\BibitemShut {NoStop}%
\bibitem [{\citenamefont {Mermin}(1990)}]{Mermin1990}%
  \BibitemOpen
  \bibfield  {author} {\bibinfo {author} {\bibfnamefont {N.~David}\
  \bibnamefont {Mermin}},\ }\bibfield  {title} {\enquote {\bibinfo {title}
  {Extreme quantum entanglement in a superposition of macroscopically distinct
  states},}\ }\href {\doibase 10.1103/PhysRevLett.65.1838} {\bibfield
  {journal} {\bibinfo  {journal} {Phys. Rev. Lett.}\ }\textbf {\bibinfo
  {volume} {65}},\ \bibinfo {pages} {1838--1840} (\bibinfo {year}
  {1990})}\BibitemShut {NoStop}%
\bibitem [{\citenamefont {Colbeck}(2009)}]{colbeck2009quantum}%
  \BibitemOpen
  \bibfield  {author} {\bibinfo {author} {\bibfnamefont {Roger}\ \bibnamefont
  {Colbeck}},\ }\bibfield  {title} {\enquote {\bibinfo {title} {Quantum and
  relativistic protocols for secure multi-party computation},}\ }\href@noop {}
  {\bibfield  {journal} {\bibinfo  {journal} {arXiv preprint arXiv:0911.3814}\
  } (\bibinfo {year} {2009})}\BibitemShut {NoStop}%
\bibitem [{\citenamefont {Colbeck}\ and\ \citenamefont
  {Kent}(2011)}]{colbeck2011private}%
  \BibitemOpen
  \bibfield  {author} {\bibinfo {author} {\bibfnamefont {Roger}\ \bibnamefont
  {Colbeck}}\ and\ \bibinfo {author} {\bibfnamefont {Adrian}\ \bibnamefont
  {Kent}},\ }\bibfield  {title} {\enquote {\bibinfo {title} {Private randomness
  expansion with untrusted devices},}\ }\href {\doibase
  10.1088/1751-8113/44/9/095305} {\bibfield  {journal} {\bibinfo  {journal}
  {Journal of Physics A: Mathematical and Theoretical}\ }\textbf {\bibinfo
  {volume} {44}},\ \bibinfo {pages} {095305} (\bibinfo {year}
  {2011})}\BibitemShut {NoStop}%
\bibitem [{\citenamefont {Kaniewski}(2016)}]{PhysRevLett.117.070402}%
  \BibitemOpen
  \bibfield  {author} {\bibinfo {author} {\bibfnamefont {J\ifmmode
  \mbox{\k{e}}\else~\k{e}\fi{}drzej}\ \bibnamefont {Kaniewski}},\ }\bibfield
  {title} {\enquote {\bibinfo {title} {Analytic and nearly optimal self-testing
  bounds for the clauser-horne-shimony-holt and mermin inequalities},}\ }\href
  {\doibase 10.1103/PhysRevLett.117.070402} {\bibfield  {journal} {\bibinfo
  {journal} {Phys. Rev. Lett.}\ }\textbf {\bibinfo {volume} {117}},\ \bibinfo
  {pages} {070402} (\bibinfo {year} {2016})}\BibitemShut {NoStop}%
\bibitem [{\citenamefont {Hall}(2016)}]{hall2016significance}%
  \BibitemOpen
  \bibfield  {author} {\bibinfo {author} {\bibfnamefont {Michael~JW}\
  \bibnamefont {Hall}},\ }\bibfield  {title} {\enquote {\bibinfo {title} {The
  significance of measurement independence for bell inequalities and
  locality},}\ }in\ \href@noop {} {\emph {\bibinfo {booktitle} {At the Frontier
  of Spacetime}}}\ (\bibinfo  {publisher} {Springer},\ \bibinfo {year} {2016})\
  pp.\ \bibinfo {pages} {189--204}\BibitemShut {NoStop}%
\bibitem [{\citenamefont {Collaboration}\ \emph {et~al.}(2018)\citenamefont
  {Collaboration} \emph {et~al.}}]{big2018challenging}%
  \BibitemOpen
  \bibfield  {author} {\bibinfo {author} {\bibfnamefont {BIG Bell~Test}\
  \bibnamefont {Collaboration}} \emph {et~al.},\ }\bibfield  {title} {\enquote
  {\bibinfo {title} {Challenging local realism with human choices},}\ }\href
  {\doibase 10.1038/s41586-018-0085-3} {\bibfield  {journal} {\bibinfo
  {journal} {Nature}\ }\textbf {\bibinfo {volume} {557}},\ \bibinfo {pages}
  {212--216} (\bibinfo {year} {2018})}\BibitemShut {NoStop}%
\bibitem [{\citenamefont {Friedman}\ \emph {et~al.}(2019)\citenamefont
  {Friedman}, \citenamefont {Guth}, \citenamefont {Hall}, \citenamefont
  {Kaiser},\ and\ \citenamefont {Gallicchio}}]{Friedman2019}%
  \BibitemOpen
  \bibfield  {author} {\bibinfo {author} {\bibfnamefont {Andrew~S.}\
  \bibnamefont {Friedman}}, \bibinfo {author} {\bibfnamefont {Alan~H.}\
  \bibnamefont {Guth}}, \bibinfo {author} {\bibfnamefont {Michael J.~W.}\
  \bibnamefont {Hall}}, \bibinfo {author} {\bibfnamefont {David~I.}\
  \bibnamefont {Kaiser}}, \ and\ \bibinfo {author} {\bibfnamefont {Jason}\
  \bibnamefont {Gallicchio}},\ }\bibfield  {title} {\enquote {\bibinfo {title}
  {Relaxed bell inequalities with arbitrary measurement dependence for each
  observer},}\ }\href {\doibase 10.1103/PhysRevA.99.012121} {\bibfield
  {journal} {\bibinfo  {journal} {Phys. Rev. A}\ }\textbf {\bibinfo {volume}
  {99}},\ \bibinfo {pages} {012121} (\bibinfo {year} {2019})}\BibitemShut
  {NoStop}%
\bibitem [{\citenamefont {Clauser}\ \emph {et~al.}(1969)\citenamefont
  {Clauser}, \citenamefont {Horne}, \citenamefont {Shimony},\ and\
  \citenamefont {Holt}}]{Clauser1969}%
  \BibitemOpen
  \bibfield  {author} {\bibinfo {author} {\bibfnamefont {John~F.}\ \bibnamefont
  {Clauser}}, \bibinfo {author} {\bibfnamefont {Michael~A.}\ \bibnamefont
  {Horne}}, \bibinfo {author} {\bibfnamefont {Abner}\ \bibnamefont {Shimony}},
  \ and\ \bibinfo {author} {\bibfnamefont {Richard~A.}\ \bibnamefont {Holt}},\
  }\bibfield  {title} {\enquote {\bibinfo {title} {Proposed experiment to test
  local hidden-variable theories},}\ }\href {\doibase
  10.1103/PhysRevLett.23.880} {\bibfield  {journal} {\bibinfo  {journal} {Phys.
  Rev. Lett.}\ }\textbf {\bibinfo {volume} {23}},\ \bibinfo {pages} {880--884}
  (\bibinfo {year} {1969})}\BibitemShut {NoStop}%
\bibitem [{\citenamefont {Mayers}\ and\ \citenamefont
  {Yao}(2003)}]{mayers2003self}%
  \BibitemOpen
  \bibfield  {author} {\bibinfo {author} {\bibfnamefont {Dominic}\ \bibnamefont
  {Mayers}}\ and\ \bibinfo {author} {\bibfnamefont {Andrew}\ \bibnamefont
  {Yao}},\ }\bibfield  {title} {\enquote {\bibinfo {title} {Self testing
  quantum apparatus},}\ }\href@noop {} {\bibfield  {journal} {\bibinfo
  {journal} {arXiv preprint quant-ph/0307205}\ } (\bibinfo {year}
  {2003})}\BibitemShut {NoStop}%
\bibitem [{\citenamefont {{Navascu{\'e}s}}\ \emph {et~al.}(2007)\citenamefont
  {{Navascu{\'e}s}}, \citenamefont {{Pironio}},\ and\ \citenamefont
  {{Ac{\'{\i}}n}}}]{NPA1}%
  \BibitemOpen
  \bibfield  {author} {\bibinfo {author} {\bibfnamefont {M.}~\bibnamefont
  {{Navascu{\'e}s}}}, \bibinfo {author} {\bibfnamefont {S.}~\bibnamefont
  {{Pironio}}}, \ and\ \bibinfo {author} {\bibfnamefont {A.}~\bibnamefont
  {{Ac{\'{\i}}n}}},\ }\bibfield  {title} {\enquote {\bibinfo {title} {{Bounding
  the Set of Quantum Correlations}},}\ }\href {\doibase
  10.1103/PhysRevLett.98.010401} {\bibfield  {journal} {\bibinfo  {journal}
  {Physical Review Letters}\ }\textbf {\bibinfo {volume} {98}},\ \bibinfo {eid}
  {010401} (\bibinfo {year} {2007})},\ \Eprint
  {http://arxiv.org/abs/quant-ph/0607119} {quant-ph/0607119} \BibitemShut
  {NoStop}%
\bibitem [{\citenamefont {Fukuda}(1997)}]{cdd}%
  \BibitemOpen
  \bibfield  {author} {\bibinfo {author} {\bibfnamefont {Komei}\ \bibnamefont
  {Fukuda}},\ }\bibfield  {title} {\enquote {\bibinfo {title} {cdd/cdd+
  reference manual},}\ }\href@noop {} {\bibfield  {journal} {\bibinfo
  {journal} {Institute for Operations Research, ETH-Zentrum}\ ,\ \bibinfo
  {pages} {91--111}} (\bibinfo {year} {1997})}\BibitemShut {NoStop}%
\bibitem [{\citenamefont {Boyd}\ and\ \citenamefont
  {Vandenberghe}(2004)}]{boyd2004convex}%
  \BibitemOpen
  \bibfield  {author} {\bibinfo {author} {\bibfnamefont {S.}~\bibnamefont
  {Boyd}}\ and\ \bibinfo {author} {\bibfnamefont {L.}~\bibnamefont
  {Vandenberghe}},\ }\href@noop {} {\emph {\bibinfo {title} {Convex
  optimization}}}\ (\bibinfo  {publisher} {Cambridge university press},\
  \bibinfo {year} {2004})\BibitemShut {NoStop}%
\bibitem [{\citenamefont {ApS}(2019)}]{mosek}%
  \BibitemOpen
  \bibfield  {author} {\bibinfo {author} {\bibfnamefont {MOSEK}\ \bibnamefont
  {ApS}},\ }\href {https://docs.mosek.com/9.0/pythonapi/index.html} {\emph
  {\bibinfo {title} {MOSEK Optimizer API for Python 9.0.104}}} (\bibinfo {year}
  {2019})\BibitemShut {NoStop}%
\bibitem [{\citenamefont {Wittek}(2015)}]{Wittek2015}%
  \BibitemOpen
  \bibfield  {author} {\bibinfo {author} {\bibfnamefont {Peter}\ \bibnamefont
  {Wittek}},\ }\bibfield  {title} {\enquote {\bibinfo {title} {Algorithm 950:
  Ncpol2sdpa—sparse semidefinite programming relaxations for polynomial
  optimization problems of noncommuting variables},}\ }\href {\doibase
  10.1145/2699464} {\bibfield  {journal} {\bibinfo  {journal} {ACM Trans. Math.
  Softw.}\ }\textbf {\bibinfo {volume} {41}},\ \bibinfo {pages} {21:1--21:12}
  (\bibinfo {year} {2015})}\BibitemShut {NoStop}%
\bibitem [{Cod()}]{Codes}%
  \BibitemOpen
  \href@noop {} {}\bibinfo {note} {Codes used for the derivation of the quantum
  bound can be found at https://github.com/RanieriVN/SecretSharing}\BibitemShut
  {NoStop}%
\end{thebibliography}%
\begin{widetext}
\appendix
\section{Simulating Mermin's inequality violation with an untrusted part}

Consider the Mermin's inequality  \cite{Mermin1990} given by
\begin{equation}
M=E_{001}+E_{010}+E_{100}-E_{111} \leq 2, 
\end{equation}
and that is violated up to its algebraic maximum $M=4$, for example, by the following non-signaling distribution:
\begin{equation}
\label{pMermin}
p(a,b,c\vert x,y,z)= \frac{1}{4} \delta_{a \oplus b \oplus c, xy \oplus xz}.
\end{equation}

To simulate this correlation with the classical causal model where Alice is the untrusted part (Fig. \ref{fig:causals}b or equivalently Fig. \ref{fig:causals}c) it is enough to consider a simple wiring using the local distribution given by
\begin{equation}
p(a,b,c\vert x,y,z)= (1/4) \delta_{a \oplus b \oplus c, y \oplus z},
\end{equation}
where $y$ and $z$ are the inputs of Bob and Charlie. Since in the model in Fig. \ref{fig:causals}b, both Bob and Charlie have access to the input $x$ of Alice, they simply have to multiply their local inputs $y$ and $z$ by $x$ and thus directly obtain the distribution \eqref{pMermin} maximally violation the Mermin inequality.

\section{Derivation of Bell inequalities with an untrusted part}

The causal structures we are working with are showed in the Fig.~\ref{fig:causals}c and Fig.~\ref{fig:causals}d.
They represent one of the two possible situations: $i)$ Alice is the untrusted part; $ii)$ Bob is the untrusted part. Since Charlie (the sender) does not know who is untrusted, our space of probabilities will be the convex combination of these two scenarios. Any distribution compatible with these two causal structures have a decomposition given by:
\begin{eqnarray}
    p_1(abce|xyz) = \sum_\lambda p(a|x\lambda)p(b|x y \lambda)p(c|xz\lambda)p(e|x\lambda)p(\lambda) \\
     p_2(abce|xyz) = \sum_\lambda p(a|x y\lambda)p(b| y \lambda)p(c|yz\lambda)p(e|y\lambda)p(\lambda).
\end{eqnarray}
As we see, each of these decompositions define a convex set, more precisely a polytope, the extremal points of which are deterministic strategies mapping inputs to outputs. For instance, $p(a|x y\lambda)= \delta_{a,f(x,y,\lambda)}$ where $f(x,y,\lambda)$ is a boolean function (that can change depending on the value of $\lambda$) mapping the inputs $x,y$ to a output value. We thus can construct the $V$-representation of the polytope (matrix of vertices corresponding to these deterministic strategies) and use a solver to find the $H$-representation (Bell inqualities) for this scenario. Notice that the Bell inequalities will only involve $p(abc|xyz)$, that is, we marginalize over $e$.

The complexity (dimensionality and number of extremal points) of this scenario is already high enough such that we cannot completely characterize the polytope in terms of the joint distributions. For this reason, we compute the extremal points (deterministic probabilities) in terms of the full correlators $(E_{xyz})$ for each causal structure and join them in one single matrix. This matrix is the $V$-representation of the our combined polytope. We then employ the python package called \textit{"cdd"}~\cite{cdd} to obtain $H$- representation, i.e., the facets of the polytope. We found $48$ facets, the non-trivial of which are symmetries of the well known Svetlichny inequality (see Eq.~\ref{snew}).

\section{Obtaining the optimal guessing probability in generalized probabilistic theories with linear programming}
Here we present how to write the maximization of the guessing probibility for the NS set as a \textit{"Linear Programming"} (LP) formulation. Our aim is to maximize the guessing probability $p(e=c|x^*y^*z^*)$ given a certain observed distribution $p(a,b,c \vert x,y,z)$ or that a certain violation of the Svetlichny inequality is observed. The code below refers to the causal structure showed in the Fig.~\ref{fig:causals}c such that Alice is the untrusted part (with similar structure and same results for the case where Bob is the untrusted part). The maximization problem can then be written as:
\begin{eqnarray}
\label{LP}
\max_{\p \in \RR^n} & & \quad \langle \vec{f}, \p \rangle   \\ \nonumber
\textrm{s.t.} & & \quad \sum_{e,a}p(abce\vert xyz) - \sum_{e,a}p(abce\vert x'yz) = 0\;\; (
\forall \; b,c,x,x',y,z) \\ \nonumber
& & \quad \sum_{b}p(abce\vert xyz) - \sum_{b}p(abce\vert xy'z) = 0 \;\;(\forall \; a,c,x,y,y',z)\\ \nonumber
& & \quad \sum_{c}p(abce\vert xyz) - \sum_{c}p(abce\vert xyz') = 0 \;\;(\forall \; a,b,x,y,z,z')\\ \nonumber
 & & \quad \sum_{a,b,c,e} \delta_{a\oplus b\oplus c,0}p(abce|x^*y^*z^*) = 1 \\ \nonumber
  & & \quad \sum_{a,b,c,e}p(abce|xyz) = 1 \;\;\;\; \forall \; x,y,z \\ \nonumber
 & & \quad \mathrm{S}\cdot\p = \gamma \\ \nonumber
 & &\quad \p \geq \0_n, \\ \nonumber  
\end{eqnarray}
where the first three equations refers to the non-signaling conditions (for Alice, Bob and Charlie), followed by the secret sharing constraint for an specific input choice $(x=x*, y=y*,z=z*)$, normalization, the violation of the Svetlichny inequality and positivity.  $\vec{f}$ represents the objective function such that $\vec{f}\cdot\p \equiv p(e=c\vert x^*y^*z^*)$.

\section{Obtaining upper bounds for the optimal guessing probability in quantum theory with semi-definite programming}

Here we tackle the problem of finding the maximal guessing probability in the scenario where Alice is the malicious party, when only a quantum channel is provided as nonlocal resource. This corresponds to finding a distribution compatible with the structure shown in Eq. \eqref{qconditions}, which in turn is compatible with the causal structure of Fig.~\ref{fig:causals}c, when the hidden variable $\Lambda$ is made quantum. Performing this optimization over all quantum behaviours is, however, a difficult problem in general \cite{Brunner2014}. To circumvent this problem, we look for upper bounds for this optimal guessing probability, retrieved by relaxations of the problem that can be solved in a easier way. 

We consider the Navascues-Pironio-Acin (NPA) hierarchy to perform this approximation \cite{NPA1}. It consists in a family of converging supersets for the set of quantum distributions, in where each level of the hierarchy is describable by a set of linear matrix inequalities in the variables of the problem. For each level of the hierarchy, the problem is then simplified to a semi-definite program (SDP) \cite{boyd2004convex, NPA1}, for which there are efficient algorithms for its solution.

Here we considered the second level of the hierarchy with extra terms combining three and four projections of distinct parties. More precisely, we considered the level $2+\textrm{ABC}+\textrm{ABE}+\textrm{BCE}+\textrm{ABCE}$ of the hierarchy, where $E$ indicates projections on Eve's state. Besides looking for a moment matrix $\mathcal{C}$ compatible with the observable distribution, we also enforce similar constraints to the ones in \eqref{LP}, pertinent to the model of Fig. \ref{fig:causals}c due to the non-signaling principle, and to the fine-tuned non-signaling structure manifest by the marginal distribution $p_{ABC}(abc\vert xyz) \coloneqq \sum_e p(abce\vert xyz)$. If we also impose that this marginal distribution equates to some fixed quantum distribution $P_{\textrm{Q}}(abc\vert xyz)$, then the problem becomes the SDP specified below:
\begin{subequations}
\label{app:SDP}
\begin{eqnarray}
\max_{p, \mathcal{C}} & & p(e=c\vert x^*y^*z^*) 
\label{app:SDP_obj}\\
\textrm{s.t.} & & \quad \sum_{a,b,c,e}p(abce|xyz) = 1 \quad\forall \; x,y,z \label{app:SDP_norm}\\
& & \quad p(abce|xyz) \geq 0 \quad\forall \; a,b,c,e,x,y,z \label{app:SDP_non} \\
& & \quad \sum_{b}p(abce\vert xyz) - \sum_{b}p(abce\vert xy'z) = 0 \quad\forall{a,c,e,x,y,z,y'} 
\label{app:SDP_NSB} \\
& & \quad \sum_{c}p(abce\vert xyz) - \sum_{c}p(abce\vert xyz') = 0 \quad\forall{a,b,e,x,y,z,z'} 
\label{app:SDP_NSC} \\
& & \quad \sum_{e,a}p(abce\vert xyz) - \sum_{e,a}p(abce\vert x'yz) = 0 \quad\forall{b,c,y,z}
\label{app:SDP_NSAE}\\
& & \quad p_{ABC}(abc\vert xyz) = P_{\textrm{Q}}(abc\vert xyz) \quad\forall{a,b,c,x,y,z}
\label{app:SDP_extra}\\
& & \quad \mathcal{C} = \mathcal{C}^\dagger,\;\; \mathcal{C} \geq 0 \label{app:SDP_NPA_triv}\\
& & \quad \sum_{jk}\Gamma_{ijk}\mathcal{C}_{jk} = \ell_i(p), \label{app:SDP_NPA_linear}
\end{eqnarray}
\end{subequations}
where the constraint \eqref{app:SDP_NPA_linear} represents sets of linear restrictions in the entries of $\mathcal{C}$ that make it compatible with a truncated moment matrix with elements $\mathcal{C}_{ij} = \mathrm{Tr}\left[M_i\,M^\dagger_j \rho \right]$, where $M_i$ are projections or sequence of projections on the parties involved. $\Gamma_{ijk}$ and $\ell_i$ are the specific coefficients and linear functions that make the correspondence between the expected value for the entry and the known data.

A solution to SDP \eqref{app:SDP} when $P_{\textrm{Q}}$ is given by
\begin{equation}
P_{\textrm{Q}} = \mathrm{Tr}\left[\left(M^a_x \otimes M^b_y \otimes M^c_z\right) \rho_v\right],
\label{app:red_prob}
\end{equation}
with $\rho_V = v\ket{GHZ}\bra{GHZ} + (1-v)\openone/8$, Alice measuring observables $\{-(\sigma_x+\sigma_y)/\sqrt{2}, (\sigma_x-\sigma_y)/\sqrt{2} \}$, and Bob and Charlie measuring $\{ \sigma_x, \sigma_y\}$, is shown as the solid curve in Fig. \ref{fig:plot2} for visibilities $0\leq v \leq 1$ of the main text.

\begin{figure}
    \centering
    \includegraphics[scale=0.5]{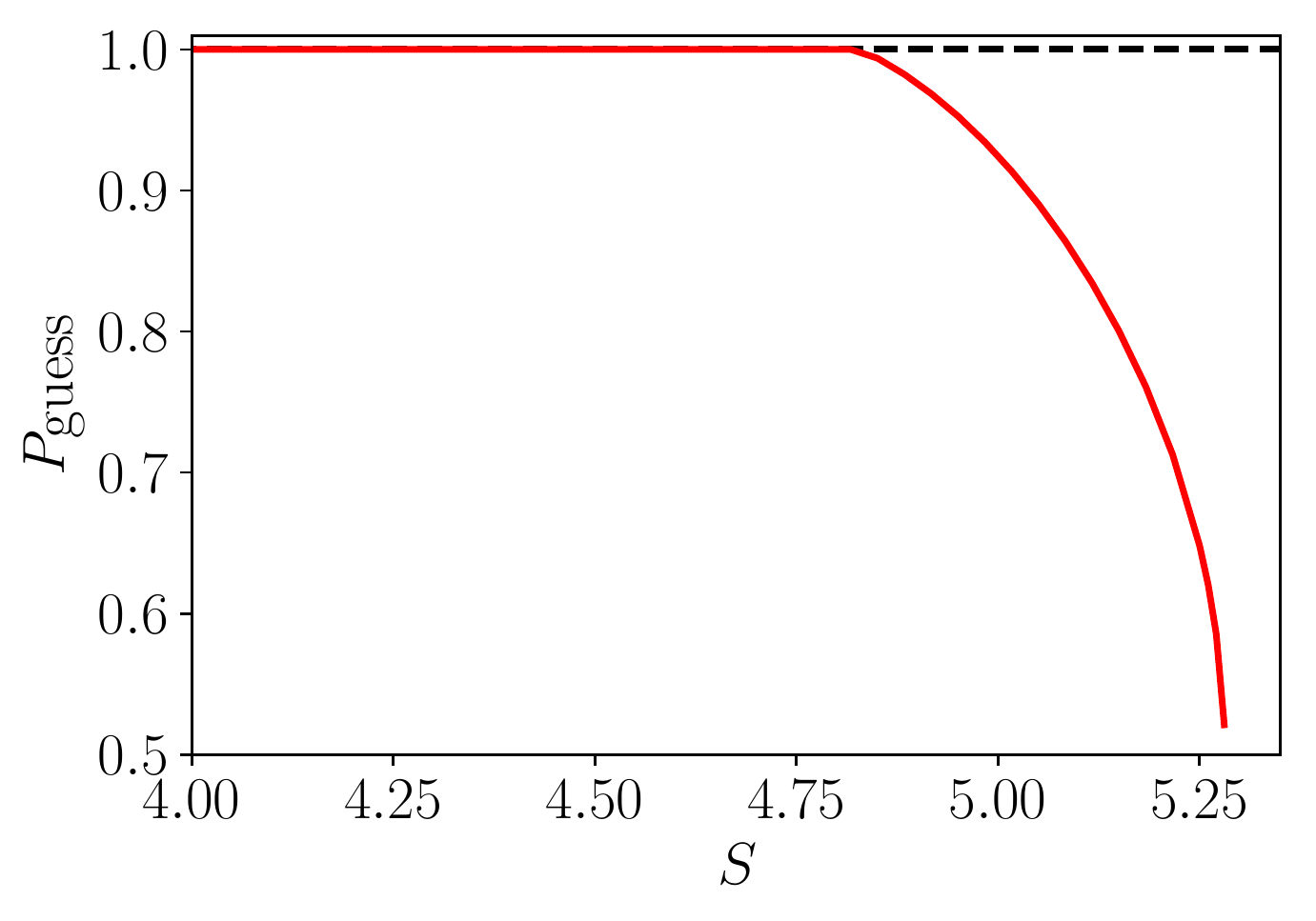}
    \caption{\textbf{Guessing probability with secret sharing condition imposed.} Upper bound for Eve's guessing probability for a generic quantum behaviour as function of the violation of inequality \eqref{snew} with the additional constraint that $a\oplus b \oplus c = 0$. It should be noted that, in contrast to the case with no secret sharing (Fig. \ref{fig:plot1} of the main text), now the optimization becomes infeasible for $S \approx 5.29$, before the quantum bound $4\sqrt{2}$. Nonetheless, it is still possible to certify some randomness on Charlie's bit as the guessing probability approaches $52\%$ for the values of violation shown here.}
    \label{app:fig_qbound_ss}
\end{figure}

Generic quantum bounds for the guessing probability in Eq.\eqref{app:SDP_obj} were obtained by replacing constraint \eqref{app:SDP_extra} for a specific violation of Svetlichny's inequality \eqref{snew} with results shown in Fig. \ref{fig:plot1} of the main text. Notice that for the quantum correlation maximally violating the Svetlichny inequality, the secret sharing condition is not satisfied. However, just as it happens in quantum key distribution protocols based on usual Bell inequality violation \cite{Ekert1991}, it is just enough that one of the parties perform a third measurement (for which the secret sharing condition $a \oplus b =c$ is satisfied).
Furthermore, this condition can also be directly imposed in the optimization problem. For instance, in Fig. \ref{app:fig_qbound_ss} we consider this same problem for specific violations of the inequality, but also ensure the secret sharing condition (without adding a third measurement input as discussed above) $a \oplus b \oplus c = 0$ when inputs are $x^*=y^*=z^*=0$ .

In all cases, to solve the SDP we used \emph{mosek} as solver \cite{mosek} and Peter Wittek's library \emph{ncpol2sdpa} \cite{Wittek2015} to model the problem and obtain the hierarchy. Codes are made available in ref \cite{Codes}.

\section{Example of an ``unsafe" NS distribution}

As discussed in the main text, below the threshold of $S=6$, the violation of the Svetlichny inequality is not enough to guarantee the security of the secret sharing protocol. 

For instance, consider the distribution
\begin{eqnarray}
p(abce\vert xyz) = \frac{1}{4}\delta_{a\oplus b\oplus c,xy\oplus xz\oplus xyz} \delta_{e\oplus c\oplus x(a\oplus b),xy\oplus xz\oplus xyz},
\label{app:unsafe_ns}
\end{eqnarray}
where $\delta_{u,v}$ is the Kronecker's delta, defined by $\delta_{u,v} = 1$ if $u=v$, and $\delta_{u,v}=0$ otherwise, and where $\oplus$ denotes addition modulo 2. It is straightforward to show that this distribution not only satisfies the required non-signaling constraints \eqref{app:SDP_NSB}-\eqref{app:SDP_NSAE}, but also violates Svetlichny's inequality by a value of $6$, guaranteeing the secret sharing condition $a\oplus b\oplus c = 0$, while at the same time providing no security to Charlie's bit, by exhibiting maximal guessing probability for Eve. 
Indeed, for $x^*=y^*=z^*=0$, the distribution reduces to $p(abce\vert x^*y^*z^*) = \delta_{a\oplus b \oplus c,0}\delta_{e\oplus c,0}/4$, which is non-zero only when both conditions $a \oplus b \oplus c=0$ and $e \oplus c = 0$ are satisfied simultaneously. While the first condition amounts to the secret sharing condition, the second one ensures that $\sum_e p(e=c\vert x^*y^*z^*) = 1$. To see that Svetlichny's inequality is violated, consider its expression as
\begin{equation}
S = \sum_{abcxyz} \beta_{abcxyz}\,P(abc\vert xyz) \leq 4,
\label{app:svet}
\end{equation}
where $\beta_{abcxyz} \coloneqq (-1)^{a+b+c+xy+xz+yz}$. For the distribution \eqref{app:unsafe_ns}, we obtain $P(abc\vert xyz) = \delta_{a\oplus b\oplus c, xy \oplus xz \oplus xyz}/4 = (1 + (-1)^{a+b+c+xy+xz+xyz})/8$ as marginal distribution for Alice, Bob and Charlie. The left-hand side (lhs) of inequality \eqref{app:svet} then reduces to $\sum_{xyz} (-1)^{(x+1)yz} = 6$, which then proves that the distribution provided is not only non-local, but also post-quantum, and yet this doesn't prevent Eve's guessing probability from reaching $1$.

If we drop the restriction to the secret sharing condition $a \oplus b \oplus c =0$, we may consider the distribution 
\begin{equation}
p^{(u)}(abce\vert xyz) = \begin{cases}
\frac{1}{16}(1 + \beta_{abc1yz}) & x=1,\cr
\frac{1}{8}\delta_{e,c} & x=0,z=0,\cr
\frac{1}{16}(1 + \beta_{abc0y1}u) & x=0,z=1,
\end{cases}
\label{app:NSdist}
\end{equation}
given in terms of the parameter $u \in [-1,\,1]$, which also satisfies the non-signaling constraints \eqref{app:SDP_NSB}-\eqref{app:SDP_NSAE} and also provides a guessing probability $p(e=c\vert x^*y^*z^*) = 1$, but that now has a tunable violation for expression \eqref{app:svet}. In fact, with $p^{(u)}$ we obtain $S = 4 + 2u$, and therefore $2 \leq S \leq 6$.

\section{Analytical proof of secret for the n participants scenario}

Now consider the case in which there are $n$ participants, one sender (which we will set as $a_1$) and $n-1$ receivers. Each part $A^{(j)}$ with two inputs $x_j=0,1$ and two outputs $a_j=0,1$, for $j=1,2,...,n$.

The $n$-partite Svetlichny inequality $S_n$ presents a recursive relation given by
\begin{eqnarray}
S_n=S_{n-1}A^{(n)}_0 + S^{\prime}_{n-1}A^{(n)}_1 \leq 4,
\end{eqnarray}
in which $S^\prime_{n-1}$ is obtained by performing the relabel $x_1\rightarrow x_1\oplus 1$ on $S_{n-1}$ \cite{Brunner2014}.

Similarly to what has been done in the main text, we can re-express the $n$-partite Svetlichny inequality as
\begin{eqnarray}
S_n & = & p_{A_n}(0\vert 0)S_{n-1}^{00} - p_{A_n}(1\vert 0)S_{n-1}^{10} + p_{A_n}(0\vert 1)S_{n-1}^{\prime 01} - p_{A_n}(1\vert 1)S_{n-1}^{\prime 1 \vert 1}.
\end{eqnarray}
In the above equation, $p_{A_n}(a_n|x_n)$ stands for the marginal distribution of part $A_n$; $S_{n-1}^{ax}$ and $S_{n-1}^{\prime ax}$ stands for the $(n-1)$-partite Svetlichny inequality for the remaining parts (conditioned on the input and output of part $A_n$).

We can now apply the same argument for $S_{n-1}^{ax}$ and $S_{n-1}^{\prime ax}$:
\begin{eqnarray}
S_n & = & p_{A_n}(0\vert 0)S_{n-1}^{00} - p_{A_n}(1 \vert 0)S_{n-1}^{10} + p_{A_n}(0\vert 1)S_{n-1}^{\prime 01} - p_{A_n}(1 \vert 1)S_{n-1}^{\prime 11}\nonumber\\
 & = & p_{A_n}(0 \vert 0)(A_{n-1}^{00}S_{n-2}^{00} + A_{n-1}^{00}S_{n-2}^{\prime 00}) -  p_{A_n}(1 \vert 0)(A_{n-1}^{10}S_{n-2}^{10} + A_{n-1}^{10}S_{n-2}^{\prime 10})\nonumber\\
 & + & p_{A_n}(0 \vert 1)(A_{n-1}^{01}S_{n-2}^{\prime 01} + A_{n-1}^{01}S_{n-2}^{01}) - p_{A_n}(1 \vert 1)(A_{n-1}^{11}S_{n-2}^{\prime 11} + A_{n-1}^{11}S_{n-2}^{11})\nonumber\\
 & = & \sum_{a_n,a_{n-1},x_n,x_{n-1}=0,1}(-1)^{a_n+a_{n-1}}p_{A_n}(a_n \vert x_n)p_{A_{n-1}}(a_{n-1}|x_{n-1}a_nx_n)S_{n-2}^{(x_n\oplus x_{n-1}) a_{n-1}a_nx_{n-1}x_n}\nonumber\\
 & = & \sum_{a_n,a_{n-1},x_n,x_{n-1}=0,1}(-1)^{a_n+a_{n-1}}p_{A_n,A_{n-1}}(a_{n-1}a_n \vert x_{n-1}x_n)S_{n-2}^{(x_n\oplus x_{n-1}) a_{n-1}a_nx_{n-1}x_n},
\end{eqnarray}
where we adopt the notation $S_{n-2}^{(0) a_{n-1}a_nx_{n-1}x_n} = S_{n-2}^{a_{n-1}a_nx_{n-1}x_n}$, and $S_{n-2}^{(1) a_{n-1}a_nx_{n-1}x_n} = S_{n-2}^{\prime a_{n-1}a_nx_{n-1}x_n}$

By repeating this process $n-4$ times, we obtain:
\begin{eqnarray}
S_n & = & \sum_{a_3,...,a_n=0,1}(-1)^{\sum_{j=3}^n a_j} p_{A_3...A_n}(a_3...a_n \vert x_3...x_n)CHSH^{(\bigoplus_{k=3}^n x_k) a_3...a_nx_3...x_n}
\end{eqnarray}
where $CHSH^{(1) a_3..a_nx_3...x_n} = CHSH^{a_3..a_nx_3...x_n}$, and $CHSH^{(1) a_3..a_nx_3...x_n} = CHSH^{\prime a_3..a_nx_3...x_n}$.

Hence, in order to obtain the maximal violation for $S_n$, we must have $CHSH^{(\bigoplus_{k=3}^n x_k) a_3...a_nx_3...x_n} = (-1)^{\sum_{j=3}^n a_j} 2\sqrt{2}$ for the quantum set and $CHSH^{(\bigoplus_{k=3}^n x_k) a_3...a_nx_3...x_n} = (-1)^{\sum_{j=3}^n a_j} 4$ in the non-signaling set. This implies that despite of the inputs and outputs of the $n-2$ untrusted parts, when maximally violating the Svetlichny's inequality the correlation between the sender and the single trusted receiver is secure, thus certifying the secret sharing protocol against the attack of the $n-2$ untrusted parts. Given the symmetry of the Svetlichny correlation, the same result follows for any combination of attackers. The observation of the maximal violation of the $n$-partite Svetlichny inequality certifies, in a device-independent manner, the security of the $n$-partite secret sharing protocol guaranteeing that all $n-1$ receivers must collaborate in order to recover the secret.
\end{widetext}
\end{document}